\documentclass[submission, Phys]{SciPost}
\usepackage{graphicx,bm}
\usepackage{amsmath}
\usepackage{amsfonts}
\usepackage{amssymb}
\usepackage{braket}
\usepackage{dsfont}
\usepackage{bm}
\usepackage{eucal}
\usepackage{mathtools}
\usepackage{comment}
\usepackage{subcaption}
\usepackage{wasysym}
\usepackage{doi}

\usepackage{array}
\usepackage{tikz}
\usetikzlibrary{arrows.meta}
\usetikzlibrary{3d} 
\usepackage{enumerate}
\usepackage{hyperref}
\usepackage{mathrsfs} 
\usepackage{amsfonts}
\usepackage{ifthen}
\usepackage{lettrine}
\usepackage{multirow}
\usepackage{float}
\usepackage{braket}
\usepackage{microtype}
\usepackage{mathrsfs} % to make mathscr way better
\usepackage{hyperref}% hyperrefing
\usepackage{xcolor}
\usepackage{scalerel}
\usepackage{bbm}% \mathbbm{1} and lowercase letters!

\usepackage{array}% tabells
\usepackage{parskip}% To make clear lines between paragraphs

\usepackage{xparse}

\definecolor{myred}{RGB}{204,102,102}

\usetikzlibrary{decorations.markings}
\usetikzlibrary{decorations.pathreplacing,calc,tikzmark}
\usetikzlibrary[arrows,backgrounds]
\usetikzlibrary{patterns,decorations.pathreplacing}
\usetikzlibrary{shapes}
\usetikzlibrary{mindmap}
\usetikzlibrary{plotmarks}

\definecolor{myolive}{RGB}{181,204,194}

\newcommand{\MapASEP}{{\cal S}}

\NewDocumentCommand{\ten}
  {m m m >{\TrimSpaces}G{1} >{\TrimSpaces}G{1} >{\TrimSpaces}G{1} >{\TrimSpaces}G{1}>{\TrimSpaces}G{0.8}}{%
  \filldraw[fill=myolive,thick,rounded corners=4pt]
    (#2,#3) rectangle (#2+0.5,#3+0.5);
  \ifnum#4=1\relax
    \draw[thick] (#2-0.25,#3+0.25)--(#2,#3+0.25);
  \fi
  \ifnum#5=1\relax
    \draw[thick] (#2+0.25,#3)--(#2+0.25,#3-0.25);
  \fi
  \ifnum#6=1\relax
    \draw[thick] (#2+0.5,#3+0.25)--(#2+0.75,#3+0.25);
  \fi
  \ifnum#7=1\relax
    \draw[thick] (#2+0.25,#3+0.5)--(#2+0.25,#3+0.75);
  \fi

  \node at (#2+0.24,#3+0.25) {\scalebox{#8}{$#1$}};
}

\NewDocumentCommand{\tenLeftArr}
  {m m m >{\TrimSpaces}G{1} >{\TrimSpaces}G{1} >{\TrimSpaces}G{1} >{\TrimSpaces}G{1}>{\TrimSpaces}G{0.8}}{%
  \filldraw[fill=myolive,thick,rounded corners=4pt]
    (#2,#3) rectangle (#2+0.5,#3+0.5);
  \ifnum#4=1\relax
    \draw[thick,<-] (#2-0.25,#3+0.25)--(#2,#3+0.25);
  \fi
  \ifnum#5=1\relax
    \draw[thick,-] (#2+0.25,#3)--(#2+0.25,#3-0.25);
  \fi
  \ifnum#6=1\relax
    \draw[thick,-] (#2+0.5,#3+0.25)--(#2+0.75,#3+0.25);
  \fi
  \ifnum#7=1\relax
    \draw[thick,->] (#2+0.25,#3+0.5)--(#2+0.25,#3+0.75);
  \fi

  \node at (#2+0.24,#3+0.25) {\scalebox{#8}{$#1$}};
}

\NewDocumentCommand{\tenRightArr}
  {m m m >{\TrimSpaces}G{1} >{\TrimSpaces}G{1} >{\TrimSpaces}G{1} >{\TrimSpaces}G{1}>{\TrimSpaces}G{0.8}}{%
  \filldraw[fill=myolive,thick,rounded corners=4pt]
    (#2,#3) rectangle (#2+0.5,#3+0.5);
  \ifnum#4=1\relax
    \draw[thick,-] (#2-0.25,#3+0.25)--(#2,#3+0.25);
  \fi
  \ifnum#5=1\relax
    \draw[thick,-] (#2+0.25,#3)--(#2+0.25,#3-0.25);
  \fi
  \ifnum#6=1\relax
    \draw[thick,->] (#2+0.5,#3+0.25)--(#2+0.75,#3+0.25);
  \fi
  \ifnum#7=1\relax
    \draw[thick,->] (#2+0.25,#3+0.5)--(#2+0.25,#3+0.75);
  \fi

  \node at (#2+0.24,#3+0.25) {\scalebox{#8}{$#1$}};
}

\NewDocumentCommand{\tenDownArr}
  {m m m >{\TrimSpaces}G{1} >{\TrimSpaces}G{1} >{\TrimSpaces}G{1} >{\TrimSpaces}G{1}>{\TrimSpaces}G{0.8}}{%
  \filldraw[fill=myolive,thick,rounded corners=4pt]
    (#2,#3) rectangle (#2+0.5,#3+0.5);
  \ifnum#4=1\relax
    \draw[thick,-] (#2-0.25,#3+0.25)--(#2,#3+0.25);
  \fi
  \ifnum#5=1\relax
    \draw[thick,->] (#2+0.25,#3)--(#2+0.25,#3-0.25);
  \fi
  \ifnum#6=1\relax
    \draw[thick,-] (#2+0.5,#3+0.25)--(#2+0.75,#3+0.25);
  \fi
  \ifnum#7=1\relax
    \draw[thick,-] (#2+0.25,#3+0.5)--(#2+0.25,#3+0.75);
  \fi

  \node at (#2+0.24,#3+0.25) {\scalebox{#8}{$#1$}};
}

\NewDocumentCommand{\tenDotted}
  {m m m >{\TrimSpaces}G{1} >{\TrimSpaces}G{1} >{\TrimSpaces}G{1} >{\TrimSpaces}G{1}>{\TrimSpaces}G{0.8}}{%
  \filldraw[fill=myolive,thick,rounded corners=4pt]
    (#2,#3) rectangle (#2+0.5,#3+0.5);

  \ifnum#4=1\relax
    \draw[thick,dotted] (#2-0.25,#3+0.25)--(#2,#3+0.25);
  \fi
  \ifnum#5=1\relax
    \draw[thick,dotted] (#2+0.25,#3)--(#2+0.25,#3-0.25);
  \fi
  \ifnum#6=1\relax
    \draw[thick,dotted] (#2+0.5,#3+0.25)--(#2+0.75,#3+0.25);
  \fi
  \ifnum#7=1\relax
    \draw[thick,dotted] (#2+0.25,#3+0.5)--(#2+0.25,#3+0.75);
  \fi

  \node at (#2+0.24,#3+0.25) {\scalebox{#8}{$#1$}};
}

\NewDocumentCommand{\tenDottedWidth}
  {m m m >{\TrimSpaces}G{1} >{\TrimSpaces}G{1} >{\TrimSpaces}G{1} >{\TrimSpaces}G{1} >{\TrimSpaces}G{0.8} >{\TrimSpaces}G{0.5}}{%
  \pgfmathsetmacro{\w}{#9} % width
  \pgfmathsetmacro{\h}{0.5} % fixed height

  \filldraw[fill=myolive,thick,rounded corners=4pt]
    (#2,#3) rectangle (#2+\w,#3+\h);

  \ifnum#4=1\relax
    \draw[thick,dotted] (#2-\w/2,#3+\h/2)--(#2,#3+\h/2);
  \fi
  \ifnum#5=1\relax
    \draw[thick,dotted] (#2+\w/2,#3)--(#2+\w/2,#3-\h/2);
  \fi
  \ifnum#6=1\relax
    \draw[thick,dotted] (#2+\w,#3+\h/2)--(#2+\w+\w/2,#3+\h/2);
  \fi
  \ifnum#7=1\relax
    \draw[thick,dotted] (#2+\w/2,#3+\h)--(#2+\w/2,#3+\h+\h/2);
  \fi

  \node at (#2+\w/2,#3+\h/2) {\scalebox{#8}{$#1$}};
}

\newcommand{\tenFour}[3]{\filldraw[fill=myolive,thick, rounded corners=4pt] (#2+0,#3+0) rectangle (#2+0.5,#3+1.5);

\draw[thick] (#2+0.5,#3+0.25)--(#2+0.75,#3+0.25);
\draw[thick] (#2-0.25,#3+0.25)--(#2,#3+0.25);
\draw[thick] (#2+0.5,#3+0.25+1)--(#2+0.75,#3+0.25+1);
\draw[thick] (#2-0.25,#3+0.25+1)--(#2,#3+0.25+1);
\node[] at (#2+0.24,#3+0.75) {\scalebox{0.8}{$#1$}}; }

\newcommand{\tenFourWidth}[4]{\filldraw[fill=myolive,thick, rounded corners=4pt] (#2+0,#3+0) rectangle (#2+0.5+#4,#3+1.5);
\draw[thick] (#2+0.5+#4,#3+0.25)--(#2+0.75+#4,#3+0.25);
\draw[thick] (#2-0.25,#3+0.25)--(#2,#3+0.25);
\draw[thick] (#2+0.5+#4,#3+0.25+1)--(#2+#4+0.75,#3+0.25+1);
\draw[thick] (#2-0.25,#3+0.25+1)--(#2,#3+0.25+1);
\node[] at (#2+0.24+0.5*#4,#3+0.75) {\scalebox{0.8}{$#1$}};}

\newcommand{\tenFourWidthHorizontal}[4]{\filldraw[fill=myolive,thick, rounded corners=4pt] (#2-0.15,#3+0.25) rectangle (#2+0.65+#4,#3+0.75);
\draw[thick] (#2,#3)--(#2,#3+0.25);
\draw[thick] (#2+#4+0.5,#3)--(#2+#4+0.5,#3+0.25);
\draw[thick] (#2,#3+0.75)--(#2,#3+0.25+0.75);
\draw[thick] (#2+#4+0.5,#3+0.75)--(#2+#4+0.5,#3+0.25+0.75);
\node[] at (#2+0.24+0.5*#4,#3+0.5) {\scalebox{0.8}{$#1$}};}

\newcommand{\tenBCS}[4]{\filldraw[fill=myred,thick, rounded corners=4pt] (#2-0.25,#3+0.25) rectangle (#2+0.25+#4,#3+0.75);
\draw[thick] (#2,#3)--(#2,#3+0.25);
\draw[thick] (#2,#3+0.75)--(#2,#3+1);
\node[] at (#2+0.5*#4,#3+0.5) {\scalebox{0.8}{$#1$}};}

\newcommand{\tenFourSpecial}[3]{\filldraw[fill=myolive,thick, rounded corners=4pt] (#2+0,#3+0) rectangle (#2+0.5,#3+1.5);
\draw[thick,dotted] (#2+0.5,#3+0.25)--(#2+0.75,#3+0.25);
\draw[thick] (#2-0.25,#3+0.25)--(#2,#3+0.25);
\draw[thick,dotted] (#2+0.5,#3+0.25+1)--(#2+0.75,#3+0.25+1);
\draw[thick] (#2-0.25,#3+0.25+1)--(#2,#3+0.25+1);
\node[] at (#2+0.24,#3+0.75) {\scalebox{0.8}{$#1$}}; }

\newcommand{\tenFourSpecialWidth}[4]{\filldraw[fill=myolive,thick, rounded corners=4pt] (#2+0,#3+0) rectangle (#2+0.5+#4,#3+1.5);
\draw[thick,dotted] (#2+0.5+#4,#3+0.25)--(#2+0.75+#4,#3+0.25);
\draw[thick] (#2-0.25,#3+0.25)--(#2,#3+0.25);
\draw[thick,dotted] (#2+0.5+#4,#3+0.25+1)--(#2+#4+0.75,#3+0.25+1);
\draw[thick] (#2-0.25,#3+0.25+1)--(#2,#3+0.25+1);
\node[] at (#2+0.24+0.5*#4,#3+0.75) {\scalebox{0.8}{$#1$}};}

\def\be{\begin{equation}}
\def\ee{\end{equation}}

\def\nn{\nonumber\\}
\def\fr#1{(\ref{#1})}

\def\re{{\rm e}}
\def\ri{{\rm i}}

\newcommand{\myDots}{\ifmmode\mathinner{\ldotp\kern-0.2em\ldotp\kern-0.2em\ldotp}\else.\kern-0.13em.\kern-0.13em.\fi}

\def\tr{{\rm Tr}}

\def\ket#1{|#1\rangle}

\usepackage{xcolor}
\input pdf-trans
\newbox\qbox
\def\usecolor#1{\csname\string\color@#1\endcsname\space}
\newcommand\bordercolor[1]{\colsplit{1}{#1}}
\newcommand\fillcolor[1]{\colsplit{0}{#1}}
\newcommand\outline[1]{\leavevmode%
  \def\maltext{\mydelim #1\mydelim}%
  \setbox\qbox=\hbox{\maltext}%
  \boxgs{Q q 2 Tr \bbthickness\space w \fillcol\space \bordercol\space}{}%
  \copy\qbox%
}
\newcommand\mathbbG[1]{\def\mydelim{$}\outline{#1}}

\newcommand\colsplit[2]{\colorlet{tmpcolor}{#2}\edef\tmp{\usecolor{tmpcolor}}%
  \def\tmpB{}\expandafter\colsplithelp\tmp\relax%
\ifnum0=#1\relax\edef\fillcol{\tmpB}\else\edef\bordercol{\tmpC}\fi}
\def\colsplithelp#1#2 #3\relax{%
  \edef\tmpB{\tmpB#1#2 }%
  \ifnum `#1>`9\relax\def\tmpC{#3}\else\colsplithelp#3\relax\fi
}
\bordercolor{black}
\fillcolor{white}
\newcommand\bbthickness{.33}
\DeclareFontFamily{U}{BOONDOX-calo}{\skewchar\font=45 }
\DeclareFontShape{U}{BOONDOX-calo}{m}{n}{
  <-> s*[1.05] BOONDOX-r-calo}{}
\DeclareFontShape{U}{BOONDOX-calo}{b}{n}{
  <-> s*[1.05] BOONDOX-b-calo}{}
\DeclareMathAlphabet{\mathcalboondox}{U}{BOONDOX-calo}{m}{n}
\SetMathAlphabet{\mathcalboondox}{bold}{U}{BOONDOX-calo}{b}{n}
\DeclareMathAlphabet{\mathbcalboondox}{U}{BOONDOX-calo}{b}{n}
\begin{document}
\begin{center}
    \Large{\bf Exact strong zero modes in quantum circuits and spin chains with non-diagonal boundary conditions}
\end{center}
	\date{\today}
	\begin{center}
	Sascha Gehrmann and Fabian H.L. Essler,
\end{center}	
\begin{center}
The Rudolf Peierls Centre for Theoretical Physics,\\ Oxford
University, Oxford OX1 3PU, UK\\
\end{center}
\section*{Abstract}
{\bf 
We construct exact strong zero mode operators (ESZM) in integrable quantum circuits and the spin-1/2 XXZ chain for general open boundary conditions, which break the bulk U(1) symmetry of the time evolution operators. We show that the ESZM is localized around one of the boundaries and induces infinite boundary coherence times. Finally we prove that the ESZM becomes spatially non-local under the map that relates the spin-1/2 XXZ chain to the asymmetric simple exclusion process, which suggests that it does not play a significant role in the dynamics of the latter.

}
	
\date{\today}
	
\tableofcontents

%%%%%%%%%%%%%%%%%%%%%%%%%%%%%%%%%%%%%%%%%%%%%%%%%%%%%%%%%%%%%%%%%%%%%%%%%%%%%%%%%%%%%%%%%%%%%%%%%%%%%%%%%%%%%%%%%%%%%%%%%%%%%%%%%%%%%%%%%%%%%%%%%%%%%%%%%%%%%%%%%%%%%%%%%%%%%%%%%%%%%%%%
\newpage

\section{Introduction}
\begin{figure}[ht]
\begin{center}
\begin{tikzpicture}

  \tikzset{
    sphere/.style={
      circle,
      minimum size=10pt, inner sep=0pt,
      shading=ball,
      ball color=blue!70!cyan!30!white, % multi-hue for depth
      draw=blue!50!black
    }
  }  
\draw[rotate around={0:(0,0)},blue] (0,0) ellipse [x radius=1, y radius=0.5];
\node[right] at (9,-0.166666666*9) {\scalebox{0.8}{$y$}};
\draw[arrows = {-Latex[width=0pt 5, length=5pt]},line width=1pt] (0,0) -- (-1*1.2,-0.75*1.2);
\node[below] at (-1.15*1.2,-0.75*1.2) {\scalebox{0.8}{$x$}};
\shade[ball color=blue!50!white, opacity=0.6] (7.5,-7.5*0.16666) circle (1);
\draw[rotate around={-2:(7.5,-7.5*0.16666666)},blue] (7.5,-7.5*0.16666) ellipse [x radius=1, y radius=0.5];
\draw[line width=1pt] (0,0) -- (6.525,-0.1666666*6.525);
\draw[dashed] (6.525,-0.1666666*6.525) -- (7.75,-0.1666666*7.75);
\draw[arrows = {-Latex[width=0pt 5, length=5pt]},line width=1pt] (8.45,-0.1666666*8.45) -- (9,-0.1666666*9);

\draw[thick,red,-{Stealth[red]}] ({0.05*cos(-160)}, {0.05*sin(-160)}) -- ({1*cos(-160)}, {0.5*sin(-160)});

\foreach \x in {0,...,5} {
  \node[sphere] at (1.5*\x, -1.5*0.166666*\x) {};
}
\foreach \x in {0,...,4} {
  \draw[line width=1pt] (1.5*\x+0.09*1.5, -1.5*0.166666*\x-1.5*0.09*0.16666) -- (1.5*\x+0.2*1.5, -1.5*0.166666*\x-1.5*0.2*0.16666) ;
}

\draw[dashed] (7.625,-0.1666666*7.625) -- (9,-0.1666666*9);  
\draw[arrows = {-Latex[width=0pt 5, length=5pt]},line width=1pt] (0,0.15) --(0,1.25*1.2);
\node[above] at (0,1.25*1.2) {\scalebox{0.8}{$z$}};
\draw[arrows = {-Latex[width=0pt 5, length=5pt]},line width=1pt] (-0.1,-0.075) -- (-1*1.2,-0.75*1.2);

\def\xPos{0.6}
\def\yPos{0.6}
\draw[thick,red,-{Stealth[red]}] (7.5+0.065,-7.5*0.16666+0.065) -- (7.5+\xPos,-7.5*0.16666+\yPos);
\draw[dotted,red]  (7.5+\xPos,-7.5*0.16666+\yPos)--(7.5+\xPos,-7.5*0.16666-0.65*\yPos);
\draw[red]  (7.5+1*0.1,-7.5*0.16666-0.5*0.1)--(7.5+\xPos,-7.5*0.16666-0.65*\yPos);
\draw[dashed] (7.5,-7.5*0.16666+0.15) -- (7.5,-7.5*0.16666+1.5);
\draw[dashed] (7.5-0.1,-7.5*0.16666-0.75*0.1) -- (7.5-1*1.,-7.5*0.16666-0.75*1.);
\draw[arrows = {-Latex[width=0pt 5, length=5pt]},line width=1pt] (7.5,-7.5*0.16666+0.9) -- (7.5,-7.5*0.16666+1.5);
\draw[arrows = {-Latex[width=0pt 5, length=5pt]},line width=1pt] (7.5-0.55,-7.5*0.16666-0.75*0.55) -- (7.5-1*1.,-7.5*0.16666-0.75*1.);
\end{tikzpicture}
\end{center}
\caption{Allowed directions for the boundary magnetic fields $\vec{h}_{1,N}$ (red vectors) in \fr{HXXZ} for an ESZM to exist. Black vectors denote the directions in spin space. The U(1) symmetry of the Hamiltonian under rotations around the z-axis is generally broken by these boundary terms.
\label{fig:bcs}}
\end{figure}
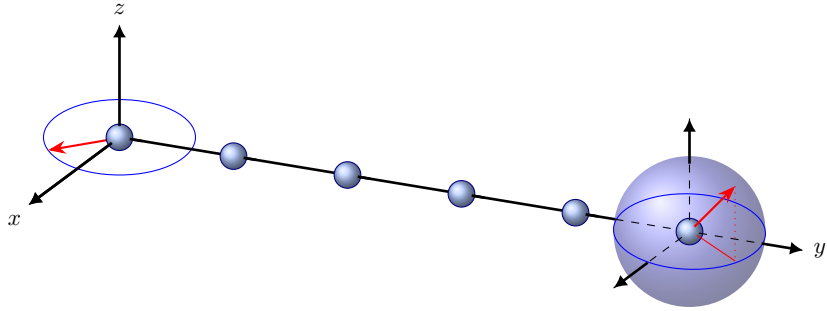

Stable edge modes in interacting many-particle systems have attracted a great deal of attention in recent years. For example, they have been known to occur in the ground state sector of models exhibiting topological order \cite{Gu2009,Pollmann2012} for some time. More recently stable, or very long-lived, edge modes have been found
at arbitrary energy densities in a range of models \cite{kitaev2001unpaired,fendley2012parafermionic,fendley2016strong,moran2017parafermionic,vasiloiu2019strong,kemp2020symmetry}. 
These strong zero mode (SZM) operators have interesting physical implications such as long coherence times of spins near the boundaries \cite{kemp2017long,else2017prethermal}. Similar edge modes can occur in periodically driven systems \cite{Sen13,Chandran14,Bahri15,iadecola2015stroboscopic,Khemani16,sreejith2016parafermion,Yao17,Potter18,mukherjee2024emergent,vernier2024strong}, stochastic processes \cite{klobas2023stochastic} and interfaces between different phases \cite{olund2023boundary}. Interestingly these modes display a certain degree of robustness under perturbations \cite{kemp2017long,yates2019almost,yates2020lifetime} even though they affect the physical behaviour at finite energy densities.
A useful definition of an SZM operator $\overline{\Psi}$ for a many-particle system with Hamiltonian $H$ is \cite{alicea2016topological} 
\begin{itemize}
\item{SZM1:}  $\| [H, \overline{\Psi}] \|={\cal O}(e^{-\alpha L})$ as $L \rightarrow\infty$.\vskip .2cm
\item{SZM2:} For some operator $D$ generating a discrete symmetry, $[\overline{\Psi}, D] \neq 0$.\vskip .2cm
\item{SZM3:} $\overline{\Psi}^n \propto \mathds{1}$ as $L\to\infty$ for some integer $n > 1$. \vskip.2cm
\end{itemize}
As was noted in Ref.~\cite{fendley2016strong}, it is possible to turn a SZM into an exact symmetry by changing the boundary conditions on one of the edges. This gives rise to an exact strong zero mode (ESZM) operator $\Psi$
\be
[\Psi,H]=0\ ,
\ee
which is localized in the vicinity of one of the boundaries. So far the analysis of SZM and ESZM operators in the literature has been restricted to boundary conditions which respect global $\mathds{Z}_2$ or $U(1)$ symmetries of the Hamiltonian. The aim of our work is to prove that this is not required for ESZM operators to exist. The picture that will emerge is most easily explained for the spin-1/2 XXZ Hamiltonian
\be
\mathbbm{H}_{\rm XXZ}=\sum^{N-1}_{j=1}\sigma^x_j\sigma^x_{j+1}+\sigma^y_j\sigma^y_{j+1}+\Delta\sigma^z_j\sigma^z_{j+1}+
\vec{h}_1\cdot\vec{\boldsymbol{\sigma}}_1+\vec{h}_N\cdot\vec{\boldsymbol{\sigma}}_N \,, \qquad\quad \Delta>1.
\label{HXXZ}
\ee
In the following we show that an ESZM operator localized around the left boundary ($j=1$) exists as long as we have
\be
h^z_1=0\ .
\label{hzcond}
\ee
This can be placed in the general SZM framework as follows. The bulk part of $\mathbbm{H}_{\rm XXZ}$ has a global $U(1)\ltimes\mathds{Z}_2$ symmetry corresponding to arbitrary rotations around the $z$-axis in spin space, and rotations $R_{\vec{n}}(\pi)$ by $\pi$ around any axis $\vec{n}$ in the $x-y$ plane. An ESZM operator $\Psi$ exists as long as the left boundary magnetic field $\vec{h}_1$ does not remove the discrete $\mathds{Z}_2$ symmetry. This requires $\vec{h}_1$ to lie in the x-y plane and the $\mathds{Z}_2$ symmetry then corresponds to $R_{\vec{h}_1}(\pi)$. The magnetic field $\vec{h}_L$ at the right boundary is allowed to be arbitrary, which is expected on physical grounds: as long as the ESZM operator is localized in the vicinity of the left boundary, the field at the far-away right boundary is not expected to affect it significantly. 
Constructing a SZM operator $\overline{\Psi}$ from $\Psi$ along the lines of \cite{fendley2016strong,essler2025strong} we then have
\be
[\overline{\Psi},R_{\vec{h}_1}(\pi)]\neq 0\ ,
\ee
thus fulfilling condition SZM2. The boundary conditions compatible with the existence of an ESZM operator are summarized in Fig.\ref{fig:bcs}. Analogous considerations apply in the periodically driven case.

The outline of this manuscript is as follows. In section \ref{sec:circuits} we construct a brick-wall quantum circuit with boundary conditions that break the $U(1)$ symmetry of the bulk circuit. The model generalizes the integrable circuit obtained by 
trotterizing the XXZ Heisenberg spin chain \cite{gritsev2017integrable,vanicat2018integrable,ljubotina2019ballistic,medenjak2020rigorous,miao2023integrable,aleiner2021bethe,claeys2022correlations,vernier2023integrable} (which corresponds to the diagonal-diagonal transfer matrix of the six-vertex model \cite{takahashi1991correlation,abraham1999correlation}) to general open boundary conditions and has the attractive feature of being readily simulable on quantum computers \cite{morvan2022formation,keenan2023evidence,maruyoshi2023conserved}. We then show that for a subset of the most general boundary conditions the model exhibits an ESZM.
In section \ref{sec:XXZ} we take the Trotter limit to establish the corresponding result for the spin-1/2 Heisenberg XXZ chain. We show that, as expected, the ESZM operator induces infinite edge coherence times for observables that have finite overlaps with the ESZM operator. In section \ref{sec:ASEP} we explore the relation between the spin-1/2 XXZ chain and the asymmetric exclusion process to investigate whether the existence of an ESZ in the former has physically significant implications for the latter. Finally, in section \ref{sec:summary} we summarize our results and comment on further developments.

%%%%%%%%%%%%%%%%%%%%%%%%%%%%%%%%%%%%%%%%%%%%%%%%%%%%%%%%
\section{Integrable brick-wall circuit with non-diagonal boundary conditions}
\label{sec:circuits}
%%%%%%%%%%%%%%%%%%%%%%%%%%%%%%%%%%%%%%%%%%%%%%%%%%%%%%%%
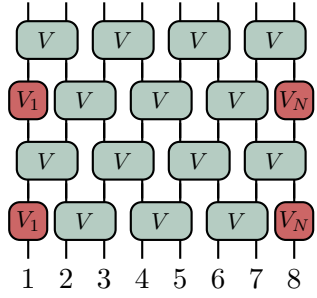
\begin{figure}[ht]
\begin{center}
\begin{tikzpicture}

\tenBCS{V_1}{1}{0}{0}
\tenBCS{V_{N}}{4.5}{0}{0}
\tenFourWidthHorizontal{V}{1.5}{0}{0}
\tenFourWidthHorizontal{V}{2.5}{0}{0}
\tenFourWidthHorizontal{V}{3.5}{0}{0}

\tenFourWidthHorizontal{V}{1}{1-0.2}{0}
\tenFourWidthHorizontal{V}{2}{1-0.2}{0}
\tenFourWidthHorizontal{V}{3}{1-0.2}{0}
\tenFourWidthHorizontal{V}{4}{1-0.2}{0}

\tenBCS{V_1}{1}{2-0.4}{0}
\tenBCS{V_{N}}{4.5}{2-0.4}{0}
\tenFourWidthHorizontal{V}{1.5}{2-0.4}{0}
\tenFourWidthHorizontal{V}{2.5}{2-0.4}{0}
\tenFourWidthHorizontal{V}{3.5}{2-0.4}{0}

\tenFourWidthHorizontal{V}{1}{3-0.6}{0}
\tenFourWidthHorizontal{V}{2}{3-0.6}{0}
\tenFourWidthHorizontal{V}{3}{3-0.6}{0}
\tenFourWidthHorizontal{V}{4}{3-0.6}{0}

\node[below] at (1,0) {$1$};
\node[below] at (1.5,0) {$2$};
\node[below] at (2,0) {$3$};
\node[below] at (2.5,0) {$4$};
\node[below] at (3,0) {$5$};
\node[below] at (3.5,0) {$6$};
\node[below] at (4,0) {$7$};
\node[below] at (4.5,0) {$8$};
\end{tikzpicture}
\end{center}
\caption{Schematic illustration of the quantum circuits analyzed in this study.
The vertical direction represents the progression of time, with the bottom (top) lines denoting the input (output) degrees of freedom. Each full cycle of evolution comprises two successive time layers: during the first, two-qubit gates act on pairs of sites $(2j,2j+1)$, and during the second, on pairs $(2j+1,2j+2)$. Red boxes represent boundary one qubit gates.  The example shown corresponds to a chain of $N=8$ qubits evolved for two cycles.\label{fig:circuit}}
\end{figure}

We start by considering a binary-drive Floquet lattice system consisting of an even number $N$ qubits. The initial state $|\psi_0\rangle$ evolves as
\be
|\psi_t\rangle=U^t|\psi_0\rangle\ ,
\ee
where the time-evolution operator takes the form of a brick-wall circuit, \emph{cf.} Fig.~\ref{fig:circuit}
\begin{equation}
\begin{aligned}
U &= U_{\mathrm{odd}}\, U_{\mathrm{even}}, \\
U_{\mathrm{even}} &= V_{1}\, V_{2,3}\, \dots \, V_{N-2,N-1}\, V_{N}, \\
U_{\mathrm{odd}} &= V_{1,2}\, V_{3,4}\, \dots \, V_{N-1,N}.
\end{aligned}
\end{equation}
Here, each \( V_{j,j+1} \) denotes a local two-qubit unitary gate acting on neighbouring sites \( (j,j+1) \), 
while single-site unitary gates \( V_{j} \) are applied at the edges. We now choose the 2-qbit gates $V_{j,j+1}$  to correspond to the integrable quantum circuit studied in Refs~\cite{gritsev2017integrable,vanicat2018integrable,ljubotina2019ballistic,medenjak2020rigorous,miao2023integrable,aleiner2021bethe,claeys2022correlations,morvan2022formation,keenan2023evidence,maruyoshi2023conserved,vernier2023integrable}
\begin{align}
V_{j,j+1}= \re^{-\frac{\ri\tau}{4}(\sigma^x_j\sigma^x_{j+1}+\sigma^y_j\sigma^y_{j+1}+\tilde{\Delta}(\sigma^z_j\sigma^z_{j+1}-\mathbbm{1}))}\,,
\label{2qbg}
\end{align}
The most general form for the single qubit gates compatible with integrability is
\begin{align}
V_1&=\re^{\ri\boldsymbol{\ell}\cdot\boldsymbol{\sigma}_1}\ ,\qquad
V_N=\re^{\ri\boldsymbol{r}\cdot\boldsymbol{\sigma}_N}\ ,
\label{1qbg}
\end{align}
where $\boldsymbol{\ell}$ and $\boldsymbol{r}$ are real vectors.
In order to ensure the existence of an ESZM we impose the following restrictions on the parameters $\tilde{\Delta}\,,\tau\in \mathbb{R}$ and $\boldsymbol{\ell}$ 
\begin{align}
\frac{\sin\big(\tfrac{\tilde{\Delta}\tau}{2}\big)}{\sin\big(\tfrac{\tau}{2}\big)}>1\,, \qquad \left|\tan\big(\tfrac{\tau}{2} \big) \sqrt{\left(\frac{\sin\big(\tfrac{\tilde{\Delta}\tau}{2}\big)}{\sin\big(\tfrac{\tau}{2}\big)}\right)^2-1}\right|\le1\, ,\quad \ell_z=0.
\label{gnjjnjnj}
\end{align}
\subsection{Integrability}
%%%%%%%%%%%%%%%%%%%%%%%%
The time evolution operator of the Floquet circuit defined above is equivalent diagonal-to-diagonal transfer matrix of the six-vertex model, which in turn can be viewed as a particular case of the inhomogeneous row-to-row transfer matrix \cite{takahashi1991correlation}. These observations allow one to bring integrability methods for spin chains with open boundaries to bear \cite{sklyanin1988boundary,de1994boundary}. The two-qubit gates can be cast in the form
\be
V_{12}=P_{1,2}R_{1,2}(\ri\delta)\ ,
\ee
where  $\delta\in[0,2\pi]$, $P_{1,2}$ is the permutation operator, and $R_{1,2}$ is the R-matrix of the XXZ model 
\cite{Korepin_book}
\begin{equation}
\begin{aligned}
R_{1,2}(u)=&\frac{\sinh(u+\tfrac{\eta}{2})}{\sinh(u+\eta)}\cosh(\tfrac{\eta}{2}) \, \sigma^0_1\,\sigma^0_2+\frac{\cosh(u+\tfrac{\eta}{2})}{\sinh(u+\eta)}\sinh(\tfrac{\eta}{2}) \, \sigma^z_1\,\sigma^z_2\\
&+\frac{\sinh(\eta)}{2\sinh(u+\eta)}\big( \sigma^x_1\sigma^x_2+\sigma^y_1\sigma^y_2 \big)\ .
\end{aligned}
\label{R12}
\end{equation}
Here $\sigma^{0}_n=\mathbbm{1}_n$ and the parameters $0<\eta$ and $\delta\in (0,\pi)\cup (\pi,2\pi)$ are related to the ones defining our circuit \fr{2qbg} by
\cite{miao2023integrable}
\begin{align}\label{dmskdmksdö}
\cosh\eta=\frac{\sin(\frac{\tilde{\Delta}\tau}{2})}{\sin(\frac{\tau}{2})}\,, \qquad \quad \sin(\delta)=-\sinh(\eta)\tan(\tfrac{\tau}{2})\,.
\end{align}
The single qubit gates \fr{1qbg} are represented in terms of so-called boundary K-matrices, which are solutions to the reflection equations \cite{cherednik1984factorizing}. The most general form appropriate for the unitary circuit \ref{fig:circuit} is
\begin{align}\label{dsnjdnsjdnsjd}
V_{1}= \tr_0\Big(K^{(L)}_0(\tfrac{\ri\delta}{2})V_{0,1}\Big)\,,\qquad\quad V_N=K^{(R)}_N(\tfrac{\ri\delta}{2})\,,
\end{align}
where \cite{de1994boundary}
\begin{align}
K^{(R)}_a(u)&=\pm\frac{1}{\sqrt{c_L}}K_a(u;\,\xi^{(R)},s^{(R)}e^{\ri\varphi^{(R)}})\ ,\quad
K^{(L)}_a(u)=\pm\frac{1}{\sqrt{c_R}}K_a(u+\eta;\xi^{(L)},s^{(L)}\re^{\ri\varphi^{(L)}})\ ,\nonumber\\
K_a(u;\xi,z)&=\sinh(2u)\big(z\sigma^+_a\pm z^*\sigma^-_a)+\sinh(u)\cosh(\xi)\sigma^z_a+\cosh(u)\sinh(\xi) \sigma^0_a .
\label{BKM}
\end{align}
\begin{comment}
\begin{align}\label{sndjsdnjsndjs}
K^{-}(u)=&\frac{1}{c^-}\begin{pmatrix}
        \sinh(u+\xi^{(R)})&s^{(R)}\,\re^{\ri \varphi^{(R)}} \sinh(2u)\\
        {\color{red}-(?)}s^{(R)}\,\re^{-\ri \varphi^{(R)}} \sinh(2u)&-\sinh(u-\xi^{(R)})
    \end{pmatrix}\ ,\nn
K^{+}(u)=&\frac{1}{c^+}\begin{pmatrix}
        \sinh(u+\eta+\xi^{(L)})& s^{(L)} \re^{\ri \varphi^{(L)}}  \sinh(2(u+\eta))\\
         - s^{(L)} \re^{-\ri \varphi^{(L)}}  \sinh(2(u+\eta))&-\sinh(u+\eta-\xi^{(L)})
    \end{pmatrix}\ .
\end{align}
Here the $u$-independent constants $c^\pm$ ensure unitarity of the time evolution
\begin{align}
    c^+=&-\frac{\cos(2\delta)-\cosh(4\eta)}{2\sin(\delta-\ri \eta)\sin(\delta+\ri \eta)}\Big(\big(s^{(L)}\big)^2 \sin^2(\delta)+\cos^2(\tfrac{\delta}{2})\Big)\,,\\
c^-=&\sqrt{|\sinh(\xi^{(R)}+i\delta/2)|^2+|s^{(R)}\sin(\delta)|^2}\ ,\nn
c^+=& \ .
\end{align}
\end{comment}
The real parameters $\boldsymbol{\ell}$ and $\boldsymbol{r}$ characterizing the quantum circuit in Eq.~\eqref{1qbg} are obtained by appropriate choices of $\varphi^{(R,L)}\in[0,2\pi)$, $s^{(R,L)}\in[0,\infty)$, and  $\xi^{(R,L)}$.  In particular, the constraint \eqref{gnjjnjnj} on $\ell^z$ translates to 
\begin{align}\label{loc_constraint}
\xi^{(L)}=\frac{\ri \pi}{2}\, ,
\end{align}
while the imaginary part of $\xi^{(R)}$ must be an integer or half integer multiple of $\pi$
in order to ensure unitary time evolution $V_1V_1^\dagger=\mathds{1}$, $V_NV_N^\dagger=\mathds{1}$.
Finally, the $\pm$ sign in \eqref{BKM} is determined by whether $\Im m(\xi^{(.)}/\pi)$ is an integer or half-integer.  The $u$-independent constants $c_{L,R}$ are normalization factors given by
\begin{equation}
\begin{aligned}
c_{L}=&-\frac{1}{2}\frac{\cos(2\delta)-\cos(4\eta)}{\sin(\delta-\ri \eta)\sin(\delta+\ri \eta)}\Big(\big(s^{(L)}\big)^2 \sin^2(\delta)+\sin(\tfrac{\delta}{2}-\ri\xi^{(L)} )\sin(\tfrac{\delta}{2}+\ri(\xi^{(L)})^* )\Big)\,,\\ 
    c_{R}=&\big(s^{(R)}\big)^2 \sin^2(\delta)+\sin\big(\tfrac{\delta}{2}-\ri \xi^{(R)}\big)\sin\big(\tfrac{\delta}{2}+\ri (\xi^{(R)})^*\big)\,.
\end{aligned}
\end{equation}

\begin{figure}[ht]
\begin{center}
\begin{tikzpicture}

\draw[thick, black] (-0.5,0) to [out=90, in=180] (-0.25,0.25);
\ten{K^{+}}{-0.75}{-0.5}{0}{0}{0}{0}{0.5};
\ten{K^{-}}{3.75}{-0.5}{0}{0}{0}{0}{0.5};

\draw[thick, black,->] (-0.5,-0.5) to [out=-90, in=180] (-0.25,-0.75);

\tenLeftArr{R^-}{0}{0}{1}{1}{1}{1};
\tenLeftArr{R^+}{1}{0}{1}{1}{1}{1};
\tenLeftArr{R^+}{3}{0}{1}{1}{1}{1};

\node[right] at (1.9,-0.25) {$\dots$};

\tenRightArr{R^+}{0}{-1}{1}{1}{1}{1};
\tenRightArr{R^-}{1}{-1}{1}{1}{1}{1};
\tenRightArr{R^-}{3}{-1}{1}{1}{1}{1};

\draw[thick, black] (3.75,-0.75) to [out=0, in=-90] (4.,-0.5);
\draw[thick, black,->] (4.,0.) to [out=90, in=0] (3.75,0.25);

\end{tikzpicture}
\end{center}
\caption{Graphical representation of the  double row-to-row transfer matrix. We used the short hand notation $R^{\pm}=R(u\pm \tfrac{\ri \delta}{2})$.}
\label{fig:TM}
\end{figure}
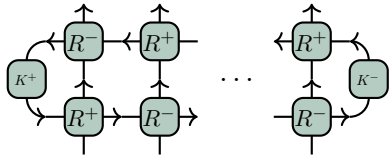

\begin{comment}
\begin{align}
|\boldsymbol{r}|&=-\frac{\ri}{2}\ln\left[\frac{\sinh(\xi^{(R)}+\frac{\ri\delta}{2})}{\sinh(\xi^{(R)}-\frac{\ri\delta}{2})}\right],\qquad
\frac{r_1-\ri r_2}{|\boldsymbol{r}|}=\frac{2s^{(R)}\re^{\ri\varphi^{(R)}}\cos(\delta/2)}{\cosh(\xi^{(R))}}\,.
\end{align}

\begin{equation}\label{fnjfdnjfn}
\begin{aligned}
\tan(t)&=2s^{(L)}\sin(\tfrac{\delta}{2})\,,\qquad p^2_z=\frac{\sec ^2\left(\frac{\delta }{2}\right) \cosh ^2(\xi^{(R)})+4 (s^{(R)})^2}{4 (s^{(R)})^2}\,,\\
\cos(pp_z)&=\frac{\sqrt{2} \cos \left(\frac{\delta }{2}\right) \sinh ( \xi^{(R)})}{\sqrt{-\cos (\delta )+\cosh (2 \xi^{(R)})+(s^{(R)})^2 (-\cos (2 \delta ))+(s^{(R)})^2}}\,.
\end{aligned}
\end{equation}
\end{comment}
The transfer matrix of the inhomogeneous six-vertex model obtained from the R-matrix \fr{R12} and boundary K-matrices \fr{BKM}
is \cite{sklyanin1988boundary}
\begin{equation}
\begin{aligned}
    \mathbbm{T}\big(u,\tfrac{\ri\delta}{2}\big)=\tr_0\Big[& K^{(L)}_0(u) R_{0,1}(u-\tfrac{\ri\delta}{2})\,R_{0,2}(u+\tfrac{\ri\delta}{2})\, \dots\,  R_{0,N}(u+\tfrac{\ri\delta}{2}) \\
    &\times K^{(R)}_0(u)\,R_{N,0}(u-\tfrac{\ri\delta}{2})\,R_{N-1,0}(u+\tfrac{\ri\delta}{2})\, \dots\, R_{1,0}(u+\tfrac{\ri\delta}{2}) \Big]\,.
\end{aligned}
\end{equation}
The transfer matrices defined in this way form a commuting family
\begin{align}
\Big[\mathbbm{T}\big(u,\tfrac{\ri\delta}{2}\big),\mathbbm{T}\big(v,\tfrac{\ri\delta}{2}\big) \Big]=0\,.
\end{align}
Expanding $\mathbbm{T}\big(v,\tfrac{\ri\delta}{2}\big)$ around $v=0$ generates a set of conserved, mutually compatible charges $\mathbbm{Q}^{(n)}$ with spatially local densities. The transfer matrix can be represented graphically as shown in Fig.~\ref{fig:TM}.
Its geometry is by construction identically to the one of our Floquet time-evolution operator and it is easy to see that the two are related by
\begin{align}
U=\mathbbm{T}\big(\tfrac{\ri\delta}{2},\tfrac{\ri\delta}{2}\big)\ .
\label{UT}
\end{align}
This in turn ensures that the conservation laws $\mathds{Q}^{(n)}$ obtained from the expansion of the transfer matrix around $u=0$ are conserved under time evolution in our Floquet circuit. Their existence precludes the time-evolving state from heating
to infinite temperature \cite{d2014long,lazarides2014equilibrium,ponte2015periodically}, and instead leads to convergence to a generalized Gibbs ensemble \cite{lazarides2014periodic,essler2016quench}.

%%%%%%%%%%%%%%%%%%%%%%%%%%%%%%%%%%%
\subsection{Exact strong zero mode operator}
%%%%%%%%%%%%%%%%%%%%%%%%%%%%%%%%%%%
We now follow Refs~\cite{vernier2023integrable,Fendley2025,essler2025strong} and construct a conserved charge from $\mathds{T}(u)$, which is spatially localized around the left boundary. As we will see, this requires us to fix the parameter $\xi^{(L)}$.
The three main steps in the construction are:
\begin{itemize}
\item{} We seek a conserved charge of the form 
\begin{align}\label{dskdnjsdnjsnsjd}
\mathbbG{\Psi}=\mathcal{N}^{-\frac{1}{2}} \mathds{T}'(u^*)\,
\end{align}
where $\mathcal{N}$ is a normalization factor that is chosen such that the Hilbert-Schmidt norm of $\mathbbG{\Psi}$ equals $1$.
\item{} The parameter $u^*$ is fixed by the requirement
\be
R_{0,n}(u^*\pm\tfrac{\ri\delta}{2})K^{(R)}_0(u^*)R_{n,0}(u^*\mp\tfrac{\ri\delta}{2})={\rm const}\ K^{(R)}_0(u^*).
\label{req}
\ee
This can be represented graphically as 
\begin{center}
\begin{tikzpicture}
\node at (2,-0.35) {$\lim\limits_{u\to u^*}$};
\ten{K}{3.75}{-0.5}{0}{0}{0}{0}{0.8};
\tenLeftArr{R^+}{3}{0}{1}{1}{1}{1};
\tenRightArr{R^-}{3}{-1}{1}{1}{1}{1};
\draw[thick, black] (3.75,-0.75) to [out=0, in=-90] (4.,-0.5);
\draw[thick, black,->] (4.,0.) to [out=90, in=0] (3.75,0.25);
\node[right] at (4.5,-0.25) {$=$};

\draw[thick, black] (3.5+2,-0.75) to [out=0, in=-90] (4.+2,-0.5);
\draw[thick, black,->] (4.+2,0.) to [out=90, in=0] (3.5+2,0.25);
\ten{K}{3.75+2}{-0.5}{0}{0}{0}{0}{0.8};
\draw[thick, ->] (6.5,-1.25) to (6.5,0.75);
\end{tikzpicture}
\end{center}
where $K=K^{(R)}_0(u^*)$ and $R^\pm=R_{0,n}(u\pm\frac{\ri\delta}{2})$.
The relation \eqref{req} imposes a particular spatial structure on $\mathbbG{\Psi}$ defined through \eqref{dskdnjsdnjsnsjd}. Applying the product rule for differentiation in \eqref{dskdnjsdnjsnsjd}, we can write $\mathbbG{\Psi}$ as a sum $\sum_j$ of terms in which the derivative acts on two $R$-matrix at site $j$ plus additional boundary contributions. For each such term the relation \eqref{req} can be applied successively from the right boundary until one reaches site $j$. This reduces each contribution to an operator that acts as the identity on sites $j+1$ to $N$, i.e. 
\be
\mathbbG{\Psi}=\sum_{j=1}^N \Psi_j\ ,
\label{psij}
\ee
where the operators $\Psi_j$ act non-trivially only on sites $1,2,\dots,j$. In our case the special value for $u^*$ is
\be
u^*=\tfrac{\ri\pi}{2}\ ,\quad K^{(R)}_0(\tfrac{\ri\pi}{2})\propto\sigma^z_0\ .
\ee
\item{} The parameters associated with the left boundary are chosen such that (see section
\ref{ssec:spatialstructure} for a derivation) 
\begin{align}
\lim_{N\to\infty}|\!|\Psi_j|\!|^2\propto e^{-\alpha j}\ ,\alpha>0\ , \qquad \text{for} \quad j\gg1,
\end{align}
where $|\!|.|\!|$ denotes the Hilbert-Schmidt norm
\begin{align}
\|\mathcal{O}\|^2=\tfrac{1}{2^N}\tr\big(\mathcal{O}^\dagger\mathcal{O}\big)\,.
\end{align}
This ensures that $\mathbbG{\Psi}$ is exponentially localized around the left boundary.  This requirement forces us to fix $\xi^{(L)}$ according to \eqref{loc_constraint}
\end{itemize}
Following Refs~\cite{vernier2023integrable,Fendley2025,essler2025strong} we can express the operator defined in this way in a convenient way using matrix-product operators (MPO) $A^\pm$
\begin{equation}\label{amkiasjnsjnajsns}
\begin{aligned}
{\mathbbG{\Psi}}=&\mathcal{N}^{-\frac{1}{2}}
B^{(L)}_{\alpha_0}\big[A^{+}\big]^{\rho_1}_{\alpha_0,\alpha_1}
[A^{-}]^{\rho_2}_{\alpha_1,\alpha_2}\dots 
[A^{+}]^{\rho_{N-1}}_{\alpha_{N-2},\alpha_{N-1}}
\big[A^{-}\big]^{\rho_N}_{\alpha_{N-1}\alpha_{N}}  B^{(R)}_{\alpha_{N}}\\
&\hskip5cm\times\, 
\sigma^{\rho_1}_1\sigma^{\rho_2}_2 ...\, \,\sigma^{\rho_N}_N\ .
\end{aligned}
\end{equation}
Here the indices $\alpha_j,\rho_k\in\{0,x,y,z\}$ are summed over.
The thermodynamic limit form of $\mathcal{N}$ is (see Appendix \ref{Normalisations} for the finite-volume expression)
\begin{align}
\mathcal{N}_{\infty}\equiv\lim_{N\to \infty}\mathcal{N}= 4 \big(1+4 \big(s^{(L)}\big)^2\big) \cosh ^2(\eta) \cosh ^2(\xi^{(R)})\,.
\end{align}
Explicit expressions for the MPOs $\big[A^\pm\big]^{\rho_j}$ and boundary vectors $B^{(R,L)}$ are given in Appendix \ref{nsjdnsjdnsjdnjs}. We remark that $[A^{\pm}]^{z}_{z0} = 0$, which is essential for ensuring the spatial locality properties of the ESZM. For $[A^{\pm}]^{z}_{z0} = 0$ to hold, the local constraint~\eqref{loc_constraint} must be satisfied. 

%%%%%%%%%%%%%%%%%%%%%%%%%%%%%%%%%%%%%%%%%%%%%%%%%%%%%%%%%%%%%%%%%%%%%%%%
\subsubsection{Spatial structure of the ESZM operator}
\label{ssec:spatialstructure}
%%%%%%%%%%%%%%%%%%%%%%%%%%%%%%%%%%%%%%%%%%%%%%%%%%%%%%%%%%%%%%%%%%%%%%%%
The representation \eqref{amkiasjnsjnajsns} allows us to obtain MPO expressions for the operators  \fr{psij}
\begin{equation}\label{sdjsggggjnn}
\begin{aligned}
\Psi_{2j-1}=&\mathcal{N}^{-\frac{1}{2}}
B^{(L)}_{\tilde{\alpha}_0}\big[A^{+}\big]^{\rho_{1}}_{\tilde{\alpha}_0,\tilde{\alpha}_1}\dots
[A^{-}]^{\rho_{2j-2}}_{\tilde{\alpha}_{2j-3},\tilde{\alpha}_{2j-2}}
[A^{+}]^{r_{2j-1}}_{\tilde{\alpha}_{2j-2},\alpha_{2j-1}}\big([A^{-}]^{0}_{\alpha_{2j-1},\alpha_{2j-1}}\big)^{\frac{N-2j}{2}}\\
&\times \big(\big[A^{+}\big]^{0}_{\alpha_{2j-1}\alpha_{2j-1}} \big)^{\frac{N-2j}{2}-1} B^{(R)}_{\alpha_{2j-1}}
\ \sigma^{\rho_1}_1\sigma^{\rho_2}_2 \dots  \sigma^{\rho_{2j-2}}_{2j-2}\sigma^{r_{2j-1}}_{2j-1}\,,\\
\Psi_{2j}=&\mathcal{N}^{-\frac{1}{2}}
B^{(L)}_{\tilde{\alpha}_0}\big[A^{+}\big]^{\rho_{1}}_{\tilde{\alpha}_0,\tilde{\alpha}_1}\dots
[A^{+}]^{\rho_{2j-1}}_{\tilde{\alpha}_{2j-2},\tilde{\alpha}_{2j-1}}
[A^{-}]^{r_{2j}}_{\tilde{\alpha}_{2j-1},\alpha_{2j}} \Big([A^{+}]^{0}_{\alpha_{2j},\alpha_{2j}}\Big)^{\frac{N-2j}{2}}\\
&\times\Big(
\big[A^{-}\big]^{0}_{\alpha_{2j}\alpha_{2j}}  \Big)^{\frac{N-2j}{2}}\, B^{(R)}_{\alpha_{2j}}\ \sigma^{\rho_1}_1\sigma^{\rho_2}_2 \dots  \sigma^{\rho_{2j-1}}_{2j-1}\sigma^{r_{2j}}_{2j}\,, 
\end{aligned}
\end{equation}
where the tilded indices $\tilde{\alpha}_\ell$ just take the values $0,x,y$, $r_{2j-1},r_{2j}\in\{x,y,z\}$ and repeated indices are summed over. We note that here the MPOs reduce to $3\times 3$ matrices due to $A^{z}_{z0}=0$.

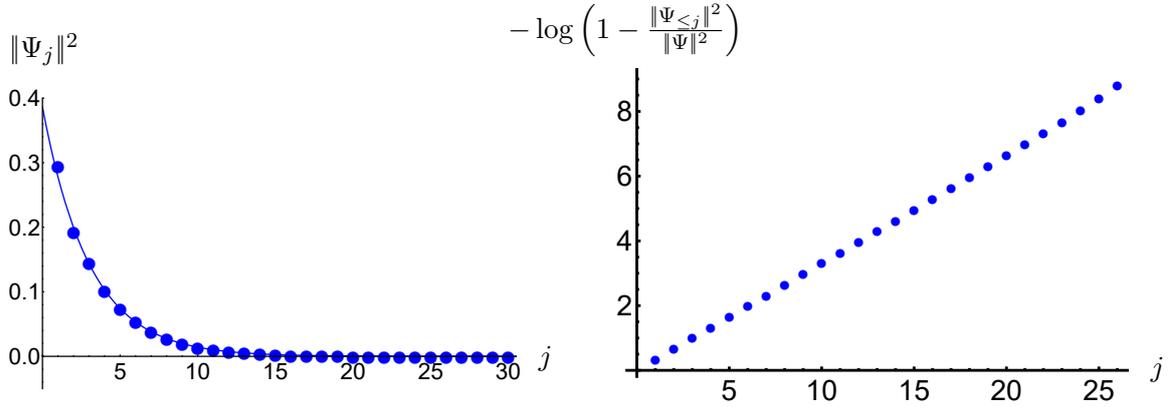
\begin{figure}[ht]
\begin{center}
\begin{tikzpicture}
\node[right] at (0,0) {
    \includegraphics[width=0.45\linewidth]{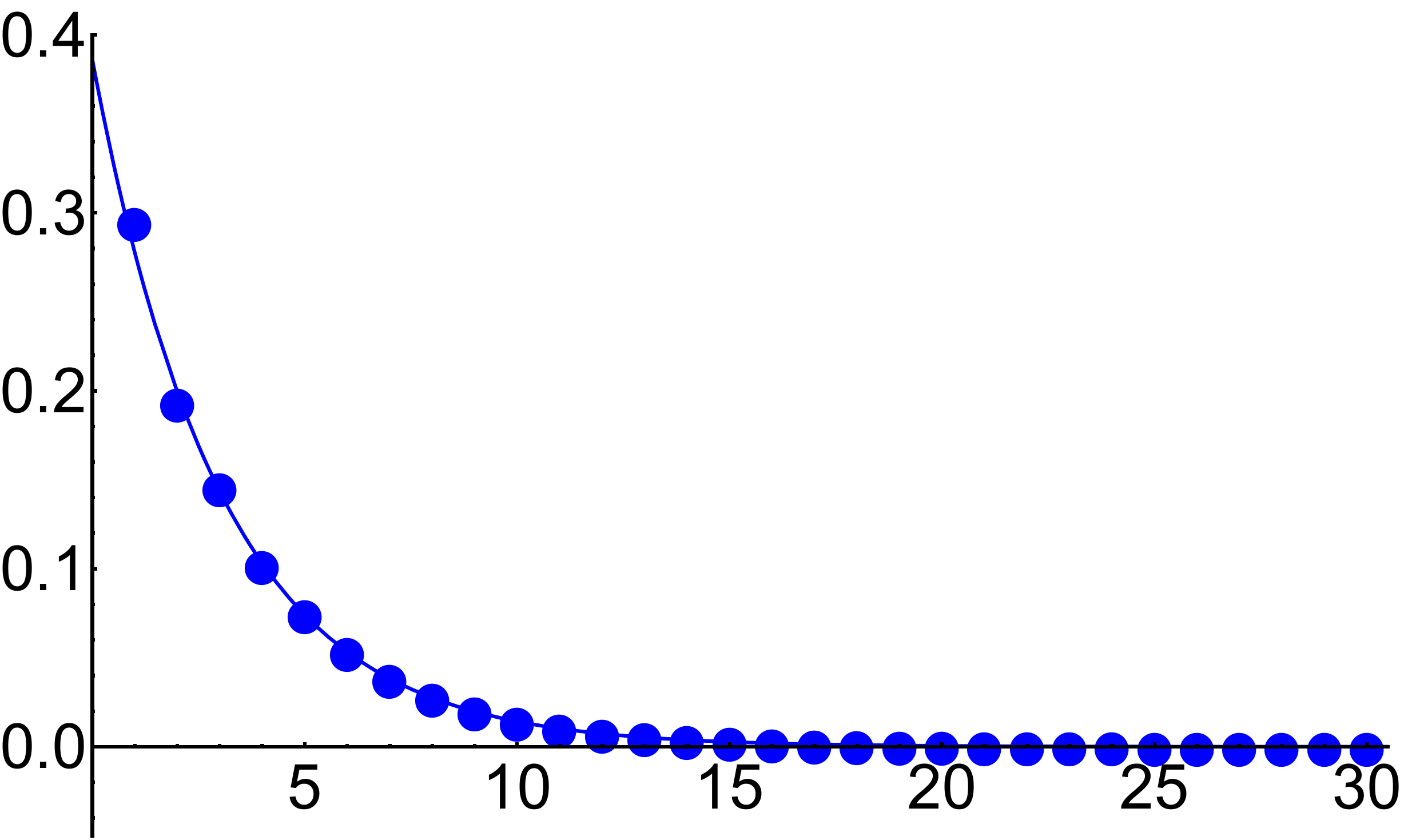}};
    \node at (0.6,2.5) {$|\!|\Psi_j|\!|^2$};
    \node at (7.2,-1.6) {$j$};
    
\node[right] at (8,0) {    \includegraphics[width=0.45\linewidth]{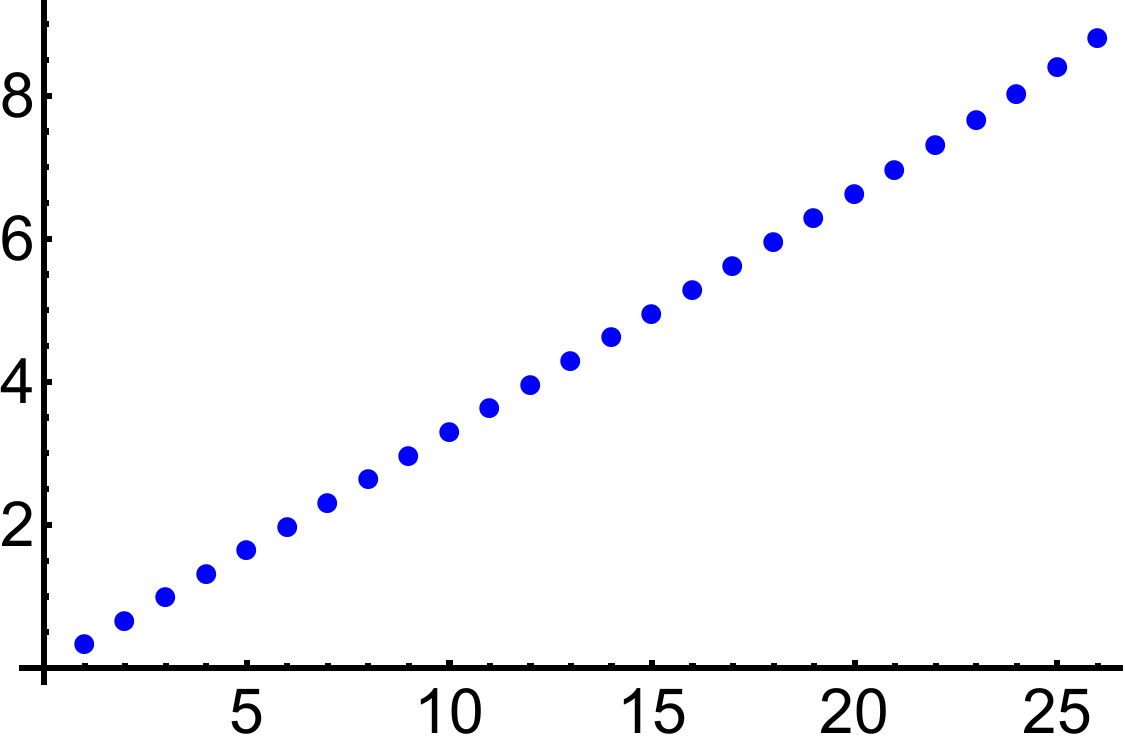}};
    \node at (8.25,2.75) {$-\log\left(1-\frac{|\!|\Psi_{\le j}|\!|^2}{|\!|\Psi|\!|^2}\right)$};
    \node at (15.25,-1.75) {$j$};
\end{tikzpicture}
\end{center}
\caption{Left: Hilbert-Schmidt Norm of $\Psi_j$ for $N=30$, $\eta=1.25$ and $\delta=0.3$ while the boundary parameters are set to be $\xi^{(R)}=0.28$, $s^{(R)}=0.32$, $s^{(L)}=0.45$, $\varphi^{(R)}=0.67$, $\varphi^{(L)}=0.97$   . Blue dots are the numerical computed data points while the blue line is the predicted asymptotic behaviour \eqref{fndnfjdnfjd}. Right: negative logarithm of residual probability. One can see clearly a linear growth. 
}
\label{fig:fnfhfhnh}
\end{figure}

As the ESZM operator is localized around the left boundary we provide explicit expressions for $\Psi_{1,2}$
\begin{equation}\label{jgkbvgjbg}
\begin{aligned}
\lim_{N\to \infty}\Psi_1=&c_{\Psi_1}  \Big (\sigma^z_1+2s^{(L)}\sin(\tfrac{\delta}{2})(\cos(\varphi^{(L)})\sigma^x_1-\sin(\varphi^{(L)})\sigma^y_1 )\Big)\,,\\
\lim_{N\to \infty}\Psi_2=&c_{\Psi_2}  \bigg(\sigma^z_2-2s^{(L)}\sin(\tfrac{\delta}{2})\cosh(\eta)\Big(\cos(\varphi^{(L)})\sigma^x_2-\sin(\varphi^{(L)})\sigma^y_2 \Big)\\
&+s^{(L)}\sinh(2\eta)\sec\big(\tfrac{\delta}{2}\big)\Big(\sin(\varphi^{(L)})\sigma^x_1\sigma^z_2+\cos(\varphi^{(L)})\sigma^y_1\sigma^z_2\Big) \\
&+2s^{(L)}\sinh(\eta)\sin\big(\tfrac{\delta}{2}\big)\tan\big(\tfrac{\delta}{2}\big)\Big(\sin(\varphi^{(L)})\sigma^z_1\sigma^x_2+\cos(\varphi^{(L)})\sigma^z_1\sigma^y_2\Big) \\
&-\sinh(\eta)\tan\big(\tfrac{\delta}{2}\big)( \sigma^x_1\sigma^y_2-\sigma^y_1\sigma^x_2)
\bigg)\ ,
\end{aligned}
\end{equation}
where
\begin{equation}\label{akfbfbfbfn}
\begin{aligned}
c_{\Psi_1} &=\mathcal{N}^{-\frac{1}{2}}_{\infty}\, \frac{2 \sinh^2(\eta)\cosh(\xi^R)\cosh(\eta)}{\ri\,\cosh(\eta-\frac{\ri \delta}{2})\cosh(\eta+\frac{\ri \delta}{2})}\ ,\qquad
c_{\Psi_2}=c_{\Psi_1} \frac{\cos^2\big(\tfrac{\delta}{2}\big)}{\cosh(\eta-\frac{\ri \delta}{2})\cosh(\eta+\frac{\ri \delta}{2})}\,.
\end{aligned}
\end{equation}
In order to prove that the ESZM operator $\mathbbG{\Psi}$ is localized around the left boundary we now consider 
the Hilbert-Schmidt norm of the $\Psi_j$   
\begin{align}
|\!|\Psi_j|\!|^2=\frac{1}{2^N}\tr\big(\Psi^\dagger_j\Psi_j\big)\,,
\end{align}
and show that these decay exponentially in $j$ (for large $j$). 
These norms can be calculated by employing the MPO representation of \eqref{sdjsggggjnn}. We have
\begin{equation}
\begin{aligned}
|\!| \Psi_{2j-1}|\!|^2&= [B^*]_{\tilde{\alpha}_0}\, [B]_{\tilde{\beta}_0}[\tilde{\mathcal{A}}]^{\tilde{\alpha}_0,\tilde{\beta}_0}_{\tilde{\alpha}_1,\tilde{\beta}_1}[\tilde{\mathcal{A}}]^{\tilde{\alpha}_1,\tilde{\beta}_1}_{\tilde{\alpha}_2,\tilde{\beta}_2}\dots [\tilde{\mathcal{A}}]^{\tilde{\alpha}_{2j-3},\tilde{\beta}_{2j-3}}_{\tilde{\alpha}_{2j-2},\tilde{\beta}_{2j-2}} [\mathcal{A}^+]^{\tilde{\alpha}_{2j-2},\tilde{\beta}_{2j-2}}_{\alpha,\beta}[C^*_{2j-1}]^\alpha\, [C_{2j-1}]^\beta\ ,\\
|\!| \Psi_{2j}|\!|^2&= [B^*]_{\tilde{\alpha}_0}\, [B]_{\tilde{\beta}_0}[\tilde{\mathcal{A}}]^{\tilde{\alpha}_0,\tilde{\beta}_0}_{\tilde{\alpha}_1,\tilde{\beta}_1}[\tilde{\mathcal{A}}]^{\tilde{\alpha}_1,\tilde{\beta}_1}_{\tilde{\alpha}_2,\tilde{\beta}_2}\dots [\tilde{\mathcal{A}}]^{\tilde{\alpha}_{2j-2},\tilde{\beta}_{2j-2}}_{\tilde{\alpha}_{2j-1},\tilde{\beta}_{2j-1}} [\mathcal{A}^{-}]^{\tilde{\alpha}_{2j-1},\tilde{\beta}_{2j-1}}_{\alpha,\beta}[C^*_{2j}]^\alpha\, [C_{2j}]^\beta\ ,\\
\end{aligned}
\end{equation}
where we have introduced the tensors
\begin{align}
[\mathcal{A}^\pm]^{\tilde{\alpha},\tilde{\beta}}_{\gamma,\zeta}&=\sum\limits_{r=x,y,z}\big[(A^\pm)^*\big]^r_{\tilde{\alpha},\gamma}\big[A^\pm\big]^r_{\tilde{\beta},\zeta}\,,\\
[\tilde{\mathcal{A}}]^{\tilde{\alpha},\tilde{\beta}}_{\tilde{\gamma},\tilde{\zeta}}&=\sum\limits_{r=0,x,y,z}\big[(A^\pm)^*\big]^r_{\tilde{\alpha},\tilde{\gamma}}\big[A^\pm\big]^r_{\tilde{\beta},\tilde{\zeta}}\,,\\
[\mathcal{C}_{2j}]^{\alpha}&=B^{(R)}_\alpha \big([A^{-}]^{0}_{\alpha,\alpha}\big)^{\frac{N-2j}{2}} \big(\big[A^{+}\big]^{0}_{\alpha,\alpha} \big)^{\frac{N-2j}{2}}\,,\\
[\mathcal{C}_{2j-1}]^{\alpha}&=B^{(R)}_\alpha \big([A^{-}]^{0}_{\alpha,\alpha}\big)^{\frac{N-2j}{2}} \big(\big[A^{+}\big]^{0}_{\alpha,\alpha} \big)^{\frac{N-2j}{2}-1}\,,\\
[B]_{\tilde{\alpha}}&=B^{(L)}_{\tilde{\alpha}}\ .
\end{align}
The expression for $\tilde{\mathcal{A}}$ turns out to be independent of the sign $\pm$ used in the RHS.
The norms can be represented graphically as
\begin{center}
\begin{tikzpicture}
\node[below] at (-1+0.3,-0.25) {\scalebox{0.8}{$1$}};
\node[below] at (0+0.5,-0.25) {\scalebox{0.8}{$2$}};
\node[below] at (1.85,-0.25) {\scalebox{0.8}{$2j-2$}};
\node[below] at (3,-0.25) {\scalebox{0.8}{$2j-1$}};
%---------------------------------Left BCs

\node[left] at (-2.25,0.75) {$|\!|\Psi_{2j-1}|\!|^2\,\,=\mathcal{N}^{-1}\cdot$};
\ten{B^*}{-2}{1}{0}{0}{1}{0}
\ten{B}{-2}{0}{0}{0}{1}{0}

\tenFourWidth{\mathcal{\tilde{A}}
}{-1}{0}{0.15}
\tenFourWidth{\mathcal{\tilde{A}}}{0+0.15}{0}{0.15}

\node[right] at (0.8,0.75) {$\dots$};
\tenFourWidth{\mathcal{\tilde{A}}}{0.5+1}{0}{0.15}
\tenFourSpecialWidth{\mathcal{A}^+}{2.65}{0}{0.15}{0.5}

\tenDottedWidth{\mathcal{C}^*_{2j-1}}{3.75}{1}{1}{0}{0}{0}{0.8}{0.8}
\tenDottedWidth{\mathcal{C}_{2j-1}}{3.75}{0}{1}{0}{0}{0}{0.8}{0.8}
%----------------------
 \pgfmathsetmacro{\offset}{2.75}
\node[below] at (-1+0.3,-0.25-\offset) {\scalebox{0.8}{$1$}};
\node[below] at (0+0.5,-0.25-\offset) {\scalebox{0.8}{$2$}};
\node[below] at (1.85,-0.25-\offset) {\scalebox{0.8}{$2j-1$}};
\node[below] at (3,-0.25-\offset) {\scalebox{0.8}{$2j$}};
%---------------------------------Left BCs

\node[left] at (-2.25,0.75-\offset)  {$|\!|\Psi_{2j}|\!|^2\,\,=\mathcal{N}^{-1}\cdot$};

\ten{B^*}{-2}{1-\offset}{0}{0}{1}{0}
\ten{B}{-2}{0-\offset}{0}{0}{1}{0}

\tenFourWidth{\mathcal{\tilde{A}}
}{-1}{0-\offset}{0.15}
\tenFourWidth{\mathcal{\tilde{A}}}{0+0.15}{0-\offset}{0.15}

\node[right] at (0.8,0.75-\offset) {$\dots$};
\tenFourWidth{\mathcal{\tilde{A}}}{0.5+1}{0-\offset}{0.15}
\tenFourSpecialWidth{\mathcal{A}^-}{2.65}{0-\offset}{0.15}{0.5}

\tenDottedWidth{\mathcal{C}^*_{2j}}{3.75}{1-\offset}{1}{0}{0}{0}{0.8}{0.5}
\tenDottedWidth{\mathcal{C}_{2j}}{3.75}{0-\offset}{1}{0}{0}{0}{0.8}{0.5}
\end{tikzpicture}
\end{center}
where the various graphical elements are defined as
\begin{center}
\begin{tikzpicture}
\tenFourSpecialWidth{\mathcal{A}^\pm}{0}{0}{0.15}
\node[right] at (-0.9,1.25) {$\tilde{\alpha}$};
\node[right] at (-0.9,0.25) {$\tilde{\beta}$};
\node[right] at (0.9,1.25) {$\gamma$};
\node[right] at (0.9,0.25) {$\zeta$};
\node[right] at (1,0.8) {$=[\mathcal{A}^\pm]^{\tilde{\alpha},\tilde{\beta}}_{\gamma,\zeta}\,,$};

\tenFourWidth{\tilde{\mathcal{A}}}{0+3.5+0.75}{0}{0.15}
\node[right] at (-0.9+3.5+0.75,1.25) {$\tilde{\alpha}$};
\node[right] at (-0.9+3.5+0.75,0.25) {$\tilde{\beta}$};
\node[right] at (0.9+3.5+0.75,1.25) {$\tilde{\gamma}$};
\node[right] at (0.9+3.5+0.75,0.25) {$\tilde{\zeta}$};
\node[right] at (1+3.5+0.75,0.8) {$=[\tilde{\mathcal{A}}]^{\tilde{\alpha},\tilde{\beta}}_{\tilde{\gamma},\tilde{\zeta}}\,,$};
\def\xx{-5}
\tenDotted{\mathcal{C}_{2j}}{7+0.2+0.5-5+\xx}{-2+0.75}{1}{0}{0}{0}{0.8}
\node[right] at (0.8+5.4-5+0.2+0.5+\xx,-1.75+0.75) {$\alpha$};
\node[right] at (0.8+7-5+0.5+\xx,-1.75+0.8){$=[\mathcal{C}_{2j}]^{\alpha}\,,$};
\def\xxx{-1}
\tenDottedWidth{\mathcal{C}_{2j-1}}{7+0.2+0.5-5-0.4+\xxx}{-3+1+0.75}{1}{0}{0}{0}{0.8}{1}
\node[right] at (0.8+5.4-5+0.2+0.5-0.6+\xxx,-2.75+0.75+1) {$\alpha$};
\node[right] at (0.8+7-5+0.5+\xxx,-2.75+0.8+1){$=[\mathcal{C}_{2j-1}]^\alpha\,,$};
\def\xxx{+3}
\ten{B}{0+2.7-0.6+\xxx}{-2.5-0.75+2}{0}{0}{1}{0}{0.8}
\node[right] at (0.8+2.7-0.6+\xxx,-2.22-0.75+2) {$\tilde{\alpha}$};
\node[right] at (1.2+2.7-0.6+\xxx,-2.20-0.75+2){$=\big[B^{(L)}\big]_{\tilde{\alpha}}\,.$};

\end{tikzpicture}
\end{center}
By grouping the indices $\tilde{\alpha},\tilde{\beta},\tilde{\gamma},\tilde{\zeta}$ into bi-indices $(\tilde{\alpha},\tilde{\beta})$, $(\tilde{\gamma},\tilde{\zeta}) $
\begin{align}
\mathcal{\tilde{A}}^{(\tilde{\alpha},\tilde{\beta})}_{(\tilde{\gamma},\tilde{\zeta})}=\mathcal{\tilde{A}}^{\tilde{\alpha},\tilde{\beta}}_{\tilde{\gamma},\tilde{\zeta}}\,,
\end{align}
we can interpret $\tilde{\mathcal{A}}$ as a matrix which can be diagonalized by means of a similarity transformation $S$:
\begin{align}\label{fmkdfjdnjfndjfndjfnj}
\mathcal{\tilde{A}}^{(\tilde{\alpha},\tilde{\beta})}_{(\tilde{\gamma},\tilde{\zeta})}=S^{(\tilde{\alpha},\tilde{\beta})}_{(\tilde{\theta},\tilde{\sigma})}\, D^{(\tilde{\theta},\tilde{\sigma})}_{(\tilde{\iota},\tilde{\rho})}\, \big[S^{-1}\big]^{(\tilde{\iota},\tilde{\rho})}_{(\tilde{\gamma},\tilde{\zeta})}\,,
\end{align}
where $D$ is a diagonal matrix. Explicit expressions for these matrices are given in appendix \ref{sakdmskdmskd}. From here on we omit the parentheses that highlight the bi-index structure
in order to ease notations.
Graphically we have
\begin{align}\label{fnsjnsdjsdjn}
\begin{tikzpicture}[baseline={(current bounding box.center)}]
\tenFour{\tilde{\mathcal{A}}}{-2}{0};
\node[right] at (-1.15,0.75) {$=$};
\tenFourWidth{S}{-0.15}{0}{0.15}
\tenFour{D}{1}{0}
\tenFourWidth{S^{-1}}{2}{0}{0.15}
\end{tikzpicture}
\end{align}
Using \eqref{fmkdfjdnjfndjfndjfnj},\eqref{fnsjnsdjsdjn}, we obtain
\begin{equation}
\begin{aligned}
|\!| \Psi^{\dagger}_{2j-1}|\!|&= [B^*]_{\tilde{\alpha}_0}\, [B]_{\tilde{\beta}_0}[S]^{\tilde{\alpha}_0,\tilde{\beta}_0}_{\tilde{\alpha}_1,\tilde{\beta}_1}[D^{2j-2}]^{\tilde{\alpha}_1,\tilde{\beta}_1}_{\tilde{\alpha}_2,\tilde{\beta}_2} [S^{-1}]^{\tilde{\alpha}_2,\tilde{\beta}_2}_{\tilde{\alpha}_{3},\tilde{\beta}_{3}} [\mathcal{A}^+]^{\tilde{\alpha}_{3},\tilde{\beta}_{3}}_{\alpha,\beta}[C^*_{2j-1}]^\alpha\, [C_{2j-1}]^\beta\ ,\\
|\!| \Psi^{\dagger}_{2j}|\!|&=[B^*]_{\tilde{\alpha}_0}\, [B]_{\tilde{\beta}_0}[S]^{\tilde{\alpha}_0,\tilde{\beta}_0}_{\tilde{\alpha}_1,\tilde{\beta}_1}[D^{2j-1}]^{\tilde{\alpha}_1,\tilde{\beta}_1}_{\tilde{\alpha}_2,\tilde{\beta}_2} [S^{-1}]^{\tilde{\alpha}_2,\tilde{\beta}_2}_{\tilde{\alpha}_{3},\tilde{\beta}_{3}} [\mathcal{A}^-]^{\tilde{\alpha}_{3},\tilde{\beta}_{3}}_{\alpha,\beta}[C^*_{2j}]^\alpha\, [C_{2j}]^\beta\ .\\
\end{aligned}
\end{equation}
These can be represented graphically as
\begin{center}
\begin{tikzpicture}
\node[left] at (-2.25,0.75) {$|\!|\Psi_{2j-1}|\!|^2\,\,=\mathcal{N}^{-1 }\cdot$};
\ten{B^*}{-2}{1}{0}{0}{1}{0}
\ten{B}{-2}{0}{0}{0}{1}{0}

\tenFourWidth{S}{-1}{0}{0.15}
\tenFourWidth{D^{2j-2}}{0+0.15}{0}{0.5}
\tenFourWidth{S^{-1}}{0.5+1}{0}{0.15}
\tenFourSpecialWidth{\mathcal{A}^+}{2.65}{0}{0.15}{0.5}

\tenDottedWidth{\mathcal{C}^*_{2j-1}}{3.75}{1}{1}{0}{0}{0}{0.8}{0.8}
\tenDottedWidth{\mathcal{C}_{2j-1}}{3.75}{0}{1}{0}{0}{0}{0.8}{0.8}
%--------------------------------------------------------------------------
\pgfmathsetmacro{\offset}{2.5}

\node[left] at (-2.25,0.75-\offset)  {$|\!|\Psi_{2j}|\!|^2\,\,=\mathcal{N}^{-1}\cdot$};
\ten{B^*}{-2}{1-\offset}{0}{0}{1}{0}
\ten{B}{-2}{0-\offset}{0}{0}{1}{0}
\tenFourWidth{S}{-1}{0-\offset}{0.15}
\tenFourWidth{D^{2j-1}}{0+0.15}{0-\offset}{0.5}
\tenFourWidth{S^{-1}}{0.5+1}{0-\offset}{0.15}
\tenFourSpecialWidth{\mathcal{A}^-}{2.65}{0-\offset}{0.15}{0.5}
\tenDottedWidth{\mathcal{C}^*_{2j}}{3.75}{1-\offset}{1}{0}{0}{0}{0.8}{0.5}
\tenDottedWidth{\mathcal{C}_{2j}}{3.75}{0-\offset}{1}{0}{0}{0}{0.8}{0.5}
\end{tikzpicture}
\end{center}
For large system sizes these expressions are dominated by the largest eigenvalue of $D$, which is smaller than $1$ provided $\eta>0$. The Hilbert-Schmidt norms therefore decay exponentially in $j$, yielding localization at the boundary for the ESZM.  The asymptotic (in $j$) rate of exponential decay in the thermodynamic limit is given by
\begin{align}\label{fndnfjdnfjd}
\lim_{N\to\infty}|\!|\Psi_j|\!|^2\asymp a_1\cdot (d_6)^{j-1} \qquad \text{for} \quad j\gg1\ ,
\end{align}
where
\begin{equation}
\begin{aligned}
d_6=&\frac{\cos^2\!\left(\frac{\delta}{2}\right)\Big(2\cos\delta-\kappa+\cosh(2\eta)\,(\kappa+1)+1\Big)}{\big(\cos\delta+\cosh(2\eta)\big)^2}<1\ ,\\
a_1=&\frac{
\sinh^2(\eta)\,\tanh^2(\eta)\,
\Big(\kappa + 4(s^{(L)})^2(\cosh(2\eta) +1) - 1\Big)\,
\Big(\sin^2\!\left(\tfrac{\delta}{2}\right)(\kappa + 1) + 2\cosh^2(\eta)\Big)
}{
(4(s^{(L)})^2 + 1)\,\kappa\,\big(\cos\delta + \cosh(2\eta)\big)^2
}\,,\\
\kappa=&\sqrt{5+4\cosh(2\eta)}\,.
\end{aligned}
\end{equation}
In Fig.~\ref{fig:fnfhfhnh} we show a comparison of the Hilbert-Schmidt norms of $\Psi_j$ for a system of $N=30$ sites with the asymptotic result \fr{fndnfjdnfjd} for a particular choice of boundary conditions. We observe that the agreement is excellent. Further, we can quantify how well a truncation of the ESZM operator to the first $j$ terms approximates the full operator. To that end we define
\be
\Psi_{\le j}=\sum_{n=1}^j\Psi_n\ ,
\ee
and then evaluate the relative difference of the square of the norms
\be
1-\frac{|\!|\Psi_{\le j}|\!|^2}{|\!|\Psi|\!|^2}\ .
\ee
As shown in Fig.~\ref{fig:fnfhfhnh}) this quantity shows an exponential decay in $j$.
%\begin{align}
%1-\frac{|\!|\Psi_{\le j}|\!|^2}{|\!|\Psi|\!|^2} \qquad \text{with} \qquad \Psi_{\le j}=\frac{1}{2^{N-j}}\tr_{j+1,\dots,N}(\mathbbG{\Psi})
%\end{align}
%which also decay exponentially (see Fig.~\ref{fig:fnfhfhnh}).

%%%%%%%%%%%%%%%%%%%%%%%%%%%%%%%%%%%%%%%%%%%%%%%%%%%%%%%%%%%%%%%%%%%%%%%%%%%%%%%%%%%%%%%%%%%%%%%%%%%%%%%%%%%%%%%%%%%%%%%%%%%%%%%%%%%%%%%%%%%%%%%%%%%%%%%%%%%%%%%%%%%%%%%%%%%%%%%%%%%%%%%%

%%%%%%%%%%%%%%%%%%%%%%%%%%%%%%%%%%%%%%%%%%%%%%%%%%%%%%%%%%%%%%%%%%%%%%%%%%%%%%%%%%%%%%%%%%%%%%%%%%%%%%%%%%%%%%%%%%%%%%%%%%%%%%%%%%%%%%%%%%%%%%%%%%%%%%%%%%%%%%%%%%%%%%%%%%%%%%%%%%%%%%%%
\section{The spin-1/2 Heisenberg XXZ model}
\label{sec:XXZ}
%%%%%%%%%%%%%%%%%%%%%%%%
A local Hamiltonian $\mathbbm{H}$ can be obtain by the first derivative of the transfer matrix at the shift point $u=0$ if we set $\delta=0$:
\begin{align}
\left.\frac{\rm d}{{\rm d}u}\right|_{u=0}\mathbbm{T}(u,0)=2\coth((\eta)\sinh(\xi^{(L)})\sinh(\xi^{(R)})\Big(\mathbbm{H}-N\cosh(\eta)+\sinh(\eta)\tanh(\eta)\Big)\,.
\end{align}
Using this normalisation, we get the standard XXZ Hamiltonian subjected to boundary magnetic field: 
\begin{equation}
\begin{aligned}
    \mathbbm{H}_{\rm XXZ}=&\sum^{N-1}_{j=1}\sigma^x_j\sigma^x_{j+1}+\sigma^y_j\sigma^y_{j+1}+\Delta\sigma^z_j\sigma^z_{j+1}+h_1+h_N \,, \qquad\quad \Delta=\cosh(\eta)\,.
\end{aligned}
\label{HXXZ}
\end{equation}
The boundary terms are given explicitly by 
\begin{equation}
\begin{aligned}
         h_1&=-2\ri s^{(L)}\sinh(\eta) \left(\re^{\ri \varphi^{(L)}}\sigma^+_1-\re^{-\ri \varphi^{(L)}} \sigma^-_1\right)\,,\\
         h_{N}&=\sinh(\eta)\coth(\xi^{(R)})\sigma^z_N+\frac{2s^{(R)}\sinh(\eta)}{\sinh(\xi^{(R)})}\left( \re^{\ri \varphi^{(R)}} \sigma^+_N+\re^{-\ri \varphi^{(R)}} \sigma^-_N\right)\,.
\end{aligned}
\end{equation}
The Hamiltonian is Hermitian, since we take $s^{(L,R)},\, \xi^{(R)} \in \mathbb{R}$ and $\varphi^{(L,R)} \in [0,2\pi]$. 
All the results discussed in the previous section carry over in complete analogy, including the existence of the SZM and its localization properties, which can be obtained by taking the limit $\delta \to 0$ of the above equations. 
This limit of the ESZM is well behaved, ensuring that no singularities arise and that the structure of the conserved operators remains intact. 
Hence, we can consistently set
\begin{align}
\lim_{\delta\to 0 }\mathbbG{\Psi} =\mathbbG{\Psi}^{\rm XXZ}\,.
\end{align}

\begin{figure}[ht]
\begin{center}
\begin{tikzpicture}
\node[right] at (0,0) {
    \includegraphics[width=0.85\linewidth]{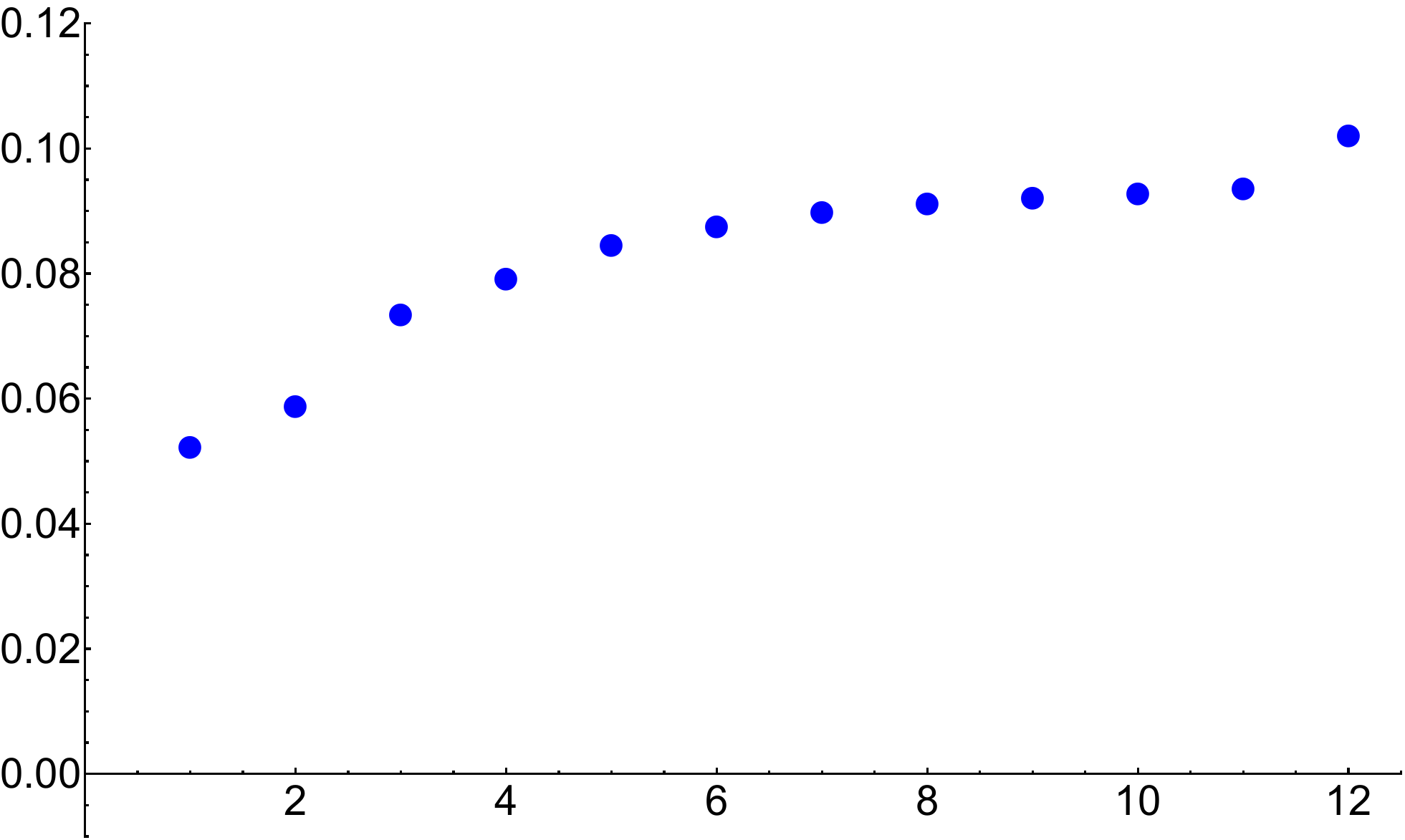}};
    \node at (1.,4.25) {$\frac{|\!|\tilde{\Psi}_j|\!|^2}{|\!|\mathbbG{\Psi}|\!|^2}$};
    \node at (13.25,-3.25) {$j$};
\end{tikzpicture}
\end{center}
\caption{Hilbert-Schmidt Norm of $\tilde{\Psi}_j$ for $N=12$, $\eta=1.37$ where the boundary parameters are  $\xi^{(L)}=1.25$, $\xi^{(R)}=0.28$, $s^{(R)}=0.12$, $s^{(L)}=0.22$, $\varphi^{(R)}=0.2$, $\varphi^{(L)}=0.5$. There is no localization close to the boundary. 
}
\label{fig:fnfhfhnjjjh}
\end{figure}

Finally we turn to the question of what happens to the ESZM in the XXZ chain when we take $\xi^{(L)}\neq \tfrac{\ri \pi}{2}$ (analogous results hold also for the quantum circuit).
 If $\xi^{(L)}\neq \tfrac{\ri \pi}{2}$ the form \eqref{amkiasjnsjnajsns} of \eqref{dskdnjsdnjsnsjd} is no longer valid. In order to assess locality properties we consider the Hilbert-Schmidt norms $|\!|\widetilde{\Psi}_j|\!|^2$, where
\begin{align}\label{fdddddkmmmss}
 \widetilde{\Psi}_j=2^{-N+j}\tr_{j+1,\dots N}({\mathbbG{\Psi}})-2^{-N+j-1}\tr_{j,\dots N}({\mathbbG{\Psi}})\, 
\end{align}
is the part of $\mathbbG{\Psi}$ that acts non-trivially on site $j$ and as the identity on sites $j+1$ to $N$.
The results of evaluating \eqref{fdddddkmmmss} for small system sizes when $\xi^{(L)}\neq\frac{\ri \pi}{2}$ 
are shown in Fig.\ref{fig:fnfhfhnjjjh} and are seen to be incompatible with localization of $\mathbbG{\Psi}$ in the vicinity of the boundary.

%%%%%%%%%%%%%%%%%%%%%%%%%%%%%%%%%%%%%%%%%%%%%%%%%%%%%%%%%%%%%%%%%%%%%%%%
\subsection{Infinite edge coherence times}
\label{sec:auto}
%%%%%%%%%%%%%%%%%%%%%%%%%%%%%%%%%%%%%%%%%%%%%%%%%%%%%%%%%%%%%%%%%%%%%%%%

A useful diagnostic for the presence of a strong zero mode are infinite temperature autocorrelation functions
\be
C^{\cal O}_0(t)=\frac{1}{2^N}{\rm Tr}\left[{\cal O}(t){\cal O}(0)\right],
\label{autocorr}
\ee
where ${\cal O}(0)$ are local operators that act non-trivially only very close to the left boundary. In generic
spin chains such correlation functions decay in time to asymptotic values that go to zero as the system size increases. 
Here the defining assumption of a \emph{generic} system is that the overlap between any conserved quantity  $\mathbbm{Q}_{n}$,  normalized as $|\!|\mathbbm{Q}_n|\!|^2=1$ for all system sizes $N$, with any operator $\mathcal{O}$ localized at the boundary (e.g. supported on a finite region adjacent to the boundary), vanishes in the thermodynamic limit $N\to \infty$.

Under this assumption, von Neumann’s mean ergodic theorem for the continuous infinite time average applied to the present case states that for any operator $\mathcal{O}$ we have
\begin{align}
\lim_{T\to \infty}\frac{1}{T} \int_0^T dt\ \mathcal{U}_t[\mathcal{O}]={\rm Proj}_{\rm C}(\mathcal{O})\ ,
\end{align}
where $\mathcal{U}_t[\mathcal{O}]=e^{iHt}\mathcal{O}e^{-iHt}$ is the Heisenberg picture time evolution operator and ${\rm Proj}_{\rm C}$ is the projection onto the eigenspace of $\mathcal{U}$ of eigenvalue $1$. The latter can be identified with the center of $\mathcal{U}$, meaning all conserved quantities.
In the generic setting defined above the overlap of any localized operator at the boundary with ${\rm Proj}_{\rm C}$ vanishes in the thermodynamic limit, so that
\begin{align}
\lim_{N\to \infty}\lim_{T\to \infty} \frac{1}{T}\int_0^Tdt\ \tr(\mathcal{U}_t[\mathcal{O}] \mathcal{O})=\lim_{N\to \infty}\tr\big[{\rm Proj}_{\rm C}(\mathcal{O})\mathcal{O}\big]=0\ .
\end{align}
The situation changes if an ESZM exists in the system, because then the overlap of ${\rm Proj}_{C}(\mathcal{O})$ with operators localized in the boundary region becomes can be finite in the thermodynamic limit. 
Let us see explicitly how an ESZM operator $\mathbbG{\Psi}$ localized around the left boundary affects $C^{\cal O}(t)$ from \eqref{autocorr}, we decompose ${\cal O}(0)$ as
\be
{\cal O}(0)=c^{\cal O}_1\mathbbG{\Psi}+c^{\cal O}_2{\cal O}'\ ,\qquad {\rm Tr}\big(\mathbbG{\Psi}^\dagger{\cal O}'\big)=0\ ,\qquad c^{\cal O}_1\neq 0\ ,
\ee
where $c^{\cal O}_1={\cal O}(L^0)$ for large system sizes and ${\cal O}'$ is by construction localized around the left boundary. Recalling that the ESZM is normalized
$|\!|\mathbbG{\Psi}|\!|^2=(2)^{-N}{\rm Tr}\left[\mathbbG{\Psi}^\dagger\mathbbG{\Psi}\right]=1$ 
then gives
\be
C^{\cal O}_0(t)=|c^{\cal O}_1|^2+|c^{\cal O}_2|^2C^{{\cal O}'}_0(t)\ .
\label{argument}
\ee
The autocorrelator $C^{{\cal O}'}(t)$ is expected to decay in time by the above argumentation for the generic setting to a value that vanishes as $N\to\infty$, which in turn implies that $C^{\cal O}(t)$ decays to a finite value $|c^{\cal O}_1|^2$ set by the overlap of ${\cal O}(0)$ with $\mathbbG{\Psi}$. In particular, we have 
\begin{align}
c^{\cal O}_1=\frac{1}{2^N} \tr(\mathbbG{\Psi}^\dagger {\cal O})\,.
\end{align}
\begin{figure}[ht]
\centering
\includegraphics[scale=0.4]{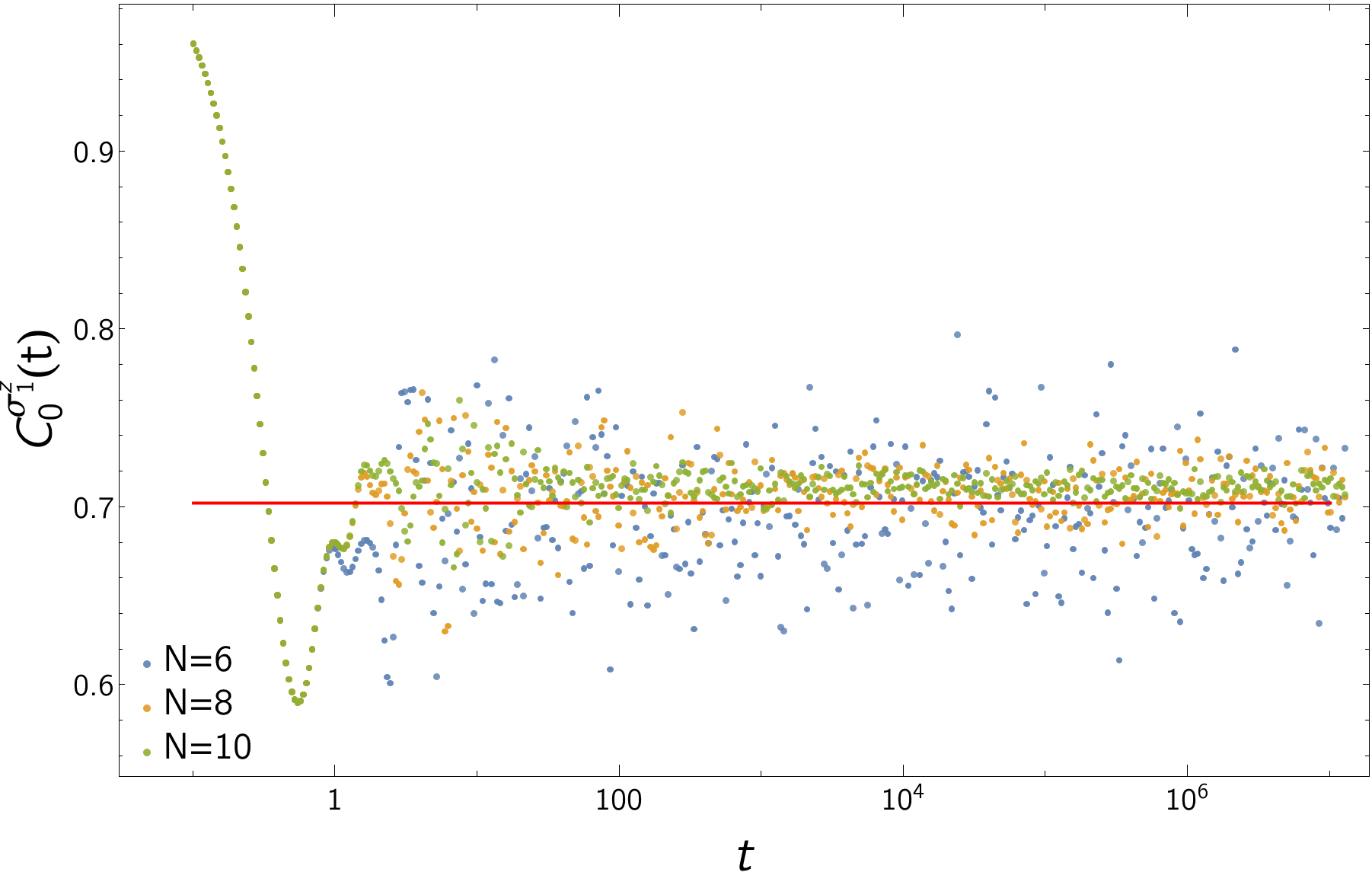}
\caption{Infinite temperature autocorrelation function for $\Delta=2.5$ and 
$h_1=(1,0.1,0)$, $h_N=(0.25,0.5,1)$. The red line is $|c_{\Psi_1}|^2$ in the thermodynamic limit.
}
\label{fig:auto1}
\end{figure}
The first few operators that are ultra-localized at the boundaries and have non-vanishing overlap with the ESZM operatopr are
\begin{align}
\sigma^z_1\,, \qquad \sigma^x_1\sigma^z_2\,,\qquad \sigma^y_1\sigma^z_2\,.
\end{align}
The asymptotic values of the infinite temperature autocorrelation function of these  in the limit $N\to \infty$ can be read off from \eqref{jgkbvgjbg}
\begin{equation}
\begin{aligned}
|c_1^{\sigma^z_1}|^2=&|c_{\Psi_1}|^2\\
|c_1^{\sigma^x_1\sigma^z_2}|^2=&|c_{\Psi_2}s^{(L)} \sinh(2\eta)\sin(\varphi^{(L)})|^2\\
|c_1^{\sigma^y_1\sigma^z_2}|^2=&|c_{\Psi_2}s^{(L)} \sinh(2\eta)\cos(\varphi^{(L)})|^2\ .
\end{aligned}
\label{limits}
\end{equation}
Here $c_{\Psi_{1,2}}$ are given by the $\delta\to0$ limit
of \eqref{akfbfbfbfn}. The edge autocorrelation function $C^{\sigma^z_1}_0(t)$ is shown for $\delta=2.5$, $N=6,8,10$ and boundary fields $h_1=(1,0.1,0)$, $h_N=(0.25,0.5,1)$ in Fig.~\ref{fig:auto1}. We observe that $C^{\sigma^z_1}_0(t)$ does not decay to zero but approaches a finite value at late times. The limiting value is in good agreement with the one predicted by
the argument presented in eqn \fr{argument} and \fr{limits}.

%%%%%%%%%%%%%%%%%%%%%%%%%%%%%%%%%%%%%%%%%%%%%%%%%%%%%%%%%%%%%%%%%%%%%%%%%%%%%%%%%%%%%%%%%%%%%%%%%%%%%%%%%%%%%%%%%%%%%%%%%%%%%%%%%%%%%%%%%%%%%%%%%%%%%%%%%%%%%%%%%%%%%%%%%%%%%%%%%%%%%%%%
\section{The asymmetric exclusion process}
\label{sec:ASEP}
%%%%%%%%%%%%%%%%%%%%%%%%%%%%%%%%%%%%%%%%%%%%%%%%%%%%%%%%%%%%%%%%%%%
There is a well-known relation between the spin-1/2 XXZ chain and the asymmetric simple exclusion process (ASEP) \cite{Gwa1992six,Gwa1992bethe,alcaraz1994reaction,sandow1994partially,essler1996representations}. The ASEP describes the biased diffusion of hard-core particles on a one-dimensional chain and is one of the best-studied paradigms of non-equilibrium Statistical Mechanics \cite{derrida1998exactly,schutz2001exactly,mallick2011some}. Of particular interest have been the effects of particle injection and extraction at the boundaries \cite{derrida1993exact,schutz1993phase,essler1996representations,sasamoto1999one,sasamoto2000density,blythe2000exact,uchiyama2004asymmetric,enaud2004large,deGier2005Bethe,de2006exact,blythe2007nonequilibrium,de2008slowest,Simon09,deGierEssler11,Lazarescu_2014}, which can lead to boundary-induced phase transitions \cite{krug1991boundary,henkel1994boundary}.

The ASEP is a stochastic process on a one-dimensional lattice with $N$ sites. At any given time $t$ each site is either occupied by a particle or empty, and the system evolves subject to the following rules. On sites $2$ to $N-1$ a particle attempts to hop one site to the right with rate $p$ and one site to the left with rate $q=1-p$. The hop is executed unless the neighbouring site is occupied, in which case the move is rejected. On the first and last sites of the lattice
these rules are modified by allowing particles to enter (leave) with rates $\alpha$ ($\gamma$) at site $i=1$ and with rates
$\delta$ ($\beta$) at site $i=N$ respectively, see Figure~\ref{fig:asep}. 
\tikzstyle{tensor}=[circle,draw=blue!50,fill=blue!20,thick]
\tikzstyle{bc}=[circle,draw=black!80,fill=black!80,thick]
\tikzstyle{wc}=[circle,draw=black!80,fill=white!80,thick]
\begin{figure}[ht]
\begin{center}
\begin{tikzpicture}[inner sep=1mm]
\draw[-] (0,0) -- (9,0);
\node[bc] at (0,0){};   
\node[bc] at (1,0){};   
\node[bc] at (3,0){};   
\node[bc] at (5,0){};
\node[bc] at (6,0){};   
\node[wc] at (2,0){};   
\node[wc] at (4,0){};
\node[wc] at (7,0){};   
\node[wc] at (8,0){};   
\node[wc] at (9,0){};
\draw[->] (5,0.25) arc[start angle=0, end angle=180, radius=0.5cm];
\draw[->] (5,0.25) arc[start angle=180, end angle=0, radius=0.5cm];
\draw[->] (9,0.25) arc[start angle=180, end angle=45, radius=0.5cm];
\draw[<-] (9,-0.25) arc[start angle=-180, end angle=-45, radius=0.5cm];
\draw[<-] (0,0.25) arc[start angle=0, end angle=135, radius=0.5cm];
\draw[->] (0,-0.25) arc[start angle=0, end angle=-130, radius=0.5cm];
\node at (4.5,1) { $q$};
\node at (5.5,1) { $p$};
\node at (9.5,1) { $\beta$};
\node at (9.5,-1) { $\delta$};
\node at (-0.5,1) { $\alpha$};
\node at (-0.5,-1) { $\gamma$};
\end{tikzpicture}
\caption{Dynamic rules for the ASEP with open boundary conditions.}
\label{fig:asep}
\end{center}
\end{figure}
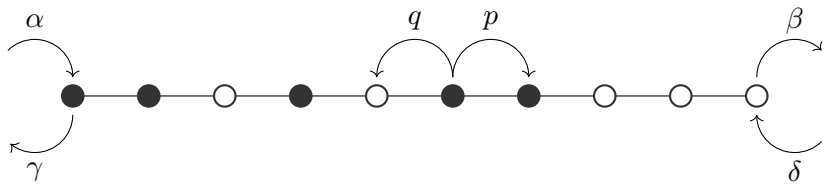
The master equation corresponding to these dynamical rules is obtained by associating a set $\boldsymbol{\tau}=\{\tau_1,\dots,\tau_N\}$ of Boolean variables with the $N$ sites of the lattice, indicating whether a particle
is present at site $i$ ($\tau_i=1$) or not ($\tau_i=0$). The state of the system at time $t$ is then characterized by the probability distribution $P_t(\tau_1,\ldots,\tau_N)$. It is convenient to associate a state in a  linear vector space with $P_t(\boldsymbol{\tau})$ by defining \cite{alcaraz1994reaction}
\begin{align}
 \ket{P_t}&=\sum_{\boldsymbol{\tau}}\, P_t(\boldsymbol{\tau})\ket{\boldsymbol{\tau}}\ ,\nonumber\\
   \ket{\boldsymbol{\tau}}&=\ket{\tau_1}\otimes\ket{\tau_2}\otimes\dots \otimes \ket{\tau_L}\,,\qquad \ket{1}=\begin{pmatrix}
        1\\0
    \end{pmatrix}\,,\qquad \ket{0}=\begin{pmatrix}
        0\\1
    \end{pmatrix}
\end{align}
The master equation then takes the form of an imaginary time Schr\"odinger equation with a non-Hermitian Hamiltonian
\begin{align}
    \frac{\partial \ket{P}}{\partial t}=\mathbbm{H}\ket{P}
\end{align}
where the Hamiltonian $\mathbbm{H}$ is given by 
\begin{align}\label{gndjnfdkfndj}
\mathbbm{H}= \sum^{N-1}_{k=1}\Big[ p\sigma^-_k\sigma^+_{k+1}+ q\sigma^+_k\sigma^-_{k+1}
+\frac{p+q}{4}(\sigma^z_j\sigma^z_{j+1}-\mathbbm{1})+\frac{p-q}{4}(\sigma^z_{k+1}-\sigma^z_{k})\Big]-B_1-B_N.
\end{align}
Here the boundary terms take the form
\begin{align}
    B_1&= \frac{\alpha-\gamma}{2}\, \sigma^z_1-\alpha \sigma^-_1-\gamma \sigma^+_1+\frac{\alpha+\gamma}{2}\ ,   \\
    B_N&= \frac{\delta-\beta}{2}\, \sigma^z_N-\delta \sigma^-_N-\beta \sigma^+_N+\frac{\beta+\delta}{2}\ .
\end{align}
Performing a similarity transformation induced by the matrix
\begin{align}
    \MapASEP=\prod^N_{j=1} \begin{pmatrix}
        1&0 \\
        0& \Lambda Q^{j-1}
    \end{pmatrix}_j
\end{align}
where $Q=\sqrt{q/p}$ and $\Lambda$ is a free parameter, we obtain the spin-1/2 XXZ chain Hamiltonian \fr{HXXZ}
\begin{align}
\MapASEP \mathbbm{H} \MapASEP^{-1}=\frac{\sqrt{pq}}{2}\big(\mathbbm{H}_{\rm XXZ}+(N-1)\cosh\eta\big)\ ,
\end{align}
where the parameters are related by \cite{essler1996representations}
\begin{align}
\cosh\eta&=\frac{p+q}{2\sqrt{pq}}\ ,\nn
h_1&=\Big(-\frac{\alpha-\gamma}{\sqrt{pq}}+ \frac{q-p}{2\sqrt{pq}}\Big) \sigma^z_1+\frac{2\Lambda \alpha}{\sqrt{pq}} \sigma^-_1+\frac{2\gamma}{\Lambda\sqrt{pq}}\sigma^+_1-\frac{\alpha+\gamma}{\sqrt{pq}}\ ,\nn
h_N&=\Big(-\frac{\delta-\beta}{\sqrt{pq}}- \frac{q-p}{2\sqrt{pq}}\Big) \sigma^z_N+\frac{2\Lambda \delta Q^{N-1}}{\sqrt{pq}} \sigma^-_N+\frac{ 2\beta Q^{1-N}}{\Lambda\sqrt{pq}} \sigma^+_N-\frac{\beta+\delta}{\sqrt{pq}}
\end{align}
The above mapping implies that the transition matrix of the ASEP commutes with the operator
\begin{align}
\mathbbG{\Psi}^{\rm ASEP}=\mathcal{N}^{\rm ASEP}\MapASEP^{-1} \mathbbG{\Psi}^{\rm XXZ}\MapASEP\ ,\qquad\quad
[\mathbbm{H},\mathbbG{\Psi}^{\rm ASEP}]=0\ ,
\end{align}
where $\mathcal{N}^{\rm ASEP}$ is a normalization factor
\begin{align}
\mathcal{N}^{\rm ASEP}&=\bigg[\prod^N_{j=2} g_j\bigg]^{-\frac{1}{2}}
\end{align}
that ensures
\begin{align}\label{dssnjfngaöfnf}
0<\lim_{N\to \infty}\frac{1}{2^N}\tr\Big(\big(\mathbbG{\Psi}^{\rm ASEP}\big)^\dagger\mathbbG{\Psi}^{\rm ASEP}\Big)<\infty\,.
\end{align}
The normalisation factor we need to choose is dependent on the regime of $\eta$. For $\eta>\log(2+\sqrt{3})$ we have
\begin{align*}
g_j=&\cosh\big((j-1)\eta+2\log(\Lambda)\big)\,,
\end{align*}
while for $\eta<\log(2+\sqrt{3})$ we need to choose 
\begin{align*}
g_j=&\cosh\big((j-1)\eta+2\log(\Lambda)\big)\frac{1}{16} \bigg(8 \text{sech}^3(\eta ) (\cosh (\eta )+\cosh (2 \eta )+\text{sech}(\eta )-1)\\&+\sinh ^2(2 \eta ) \sqrt{10 \cosh (\eta )+4 \cosh (2 \eta )+2} \text{csch}\left(\frac{\eta }{2}\right) \text{sech}^6(\eta )\bigg) \,.
\end{align*}
In order to ascertain whether the operator $\mathbbG{\Psi}^{\rm ASEP}$ is still localized around the left boundary we need to determine its spatial locality structure. The composition of the ESZM in the XXZ chain
\begin{align}
\mathbbG{\Psi}^{\rm XXZ}=\sum^N_{j=0} \Psi^{\rm XXZ}_j\,
\end{align}
is inherited by the ESZM in the ASEP
\begin{align}
\mathbbG{\Psi}^{\rm ASEP}= \mathcal{N}^{\rm ASEP}\sum^N_{j=0}\MapASEP^{-1} \Psi^{\rm XXZ}_j \MapASEP=\sum^N_{j=0}\Psi_j^{\rm ASEP}.
\label{psiasep}
\end{align}
In order to ascertain whether the $\Psi^{\rm ASEP}_j$ have good spatial locality structures we calculate their Hilbert-Schmidt norms. Using that $\MapASEP=\MapASEP^\dagger$ we have
\begin{align*}
|\!|\Psi^{\rm ASEP}_j|\!|^2
=\frac{\left(\mathcal{N}^{\rm ASEP}\right)^2}{2^N}\cdot\tr\bigg(\big(\MapASEP (\Psi^{\rm XXZ}_j)^\dagger \MapASEP^{-1} \MapASEP^{-1} \Psi^{\rm XXZ}_j \MapASEP\bigg)\ .
\end{align*}
This is conveniently evaluated using a MPO representation
\begin{center}
\begin{tikzpicture}
\node[left] at (-2.25,0.75)  {$|\!|\Psi^{\rm ASEP}_{j}|\!|^2\,\,=\mathcal{N}^{-1}\cdot$};
\ten{B^*}{-2}{1}{0}{0}{1}{0}
\ten{B}{-2}{0}{0}{0}{1}{0}
\tenFourWidth{\mathcal{A}_1
}{-1}{0}{0.15}
\tenFourWidth{\mathcal{A}_2}{0+0.15}{0}{0.15}
\node[right] at (0.78,0.75) {$\dots$};
\tenFourWidth{\mathcal{A}_{j-1}}{0.5+1}{0}{0.3}
\tenFourSpecialWidth{\mathcal{A}_{j}}{2.65}{0}{0.15}{0.5}
\tenDottedWidth{\mathcal{C}^*_{j}}{3.75}{1}{1}{0}{0}{0}{0.8}{0.5}
\tenDottedWidth{\mathcal{C}_{j}}{3.75}{0}{1}{0}{0}{0}{0.8}{0.5}
\end{tikzpicture}
\end{center}

Here the various elements are given by

\begin{tikzpicture}
\tenFour{\mathcal{A}_j}{0+7}{0}
\node[right] at (-0.8+7,1.25) {$\tilde{\alpha}$};
\node[right] at (-0.8+7,0.25) {$\tilde{\beta}$};
\node[right] at (0.8+7,1.25) {$\tilde{\gamma}$};
\node[right] at (0.8+7,0.25) {$\tilde{\delta}$};
\node[right] at (1+7,0.75) {$=\sum\limits_{\rho_1,\rho_2=0,x,y}\big[A^*\big]^{\rho_1}_{\tilde{\alpha},\tilde{\gamma}}\big[A\big]^{\rho_2}_{\tilde{\beta},\tilde{\delta}} \cdot (g_j)^{-1}\tr\big(\MapASEP_j\sigma^{\rho_1}(\MapASEP_j\MapASEP_j)^{-1}\sigma^{\rho_2}\MapASEP_j\big)$};
\end{tikzpicture}

\begin{tikzpicture}
\tenFourSpecial{\mathcal{A}_j}{0}{0}
\node[right] at (-0.8,1.25) {$\tilde{\alpha}$};
\node[right] at (-0.8,0.25) {$\tilde{\beta}$};
\node[right] at (0.8,1.25) {$\gamma$};
\node[right] at (0.8,0.25) {$\delta$};
\node[right] at (1,0.75) {$=\sum\limits_{\rho_1,\rho_2=x,y,z}\big[A^{\rho_1}_{\tilde{\alpha},\gamma}\big]^*\big[A^{\rho_2}_{\tilde{\beta},\delta}\big]\cdot (g_j)^{-1}\tr\big(\MapASEP_j\sigma^{\rho_1}(\MapASEP_j\MapASEP_j)^{-1}\sigma^{\rho_2}\MapASEP_j\big)$};
\end{tikzpicture}

\begin{tikzpicture}
\ten{B}{0+2.7-0.6}{-2.5-0.8}{0}{0}{1}{0}{0.8}
\node[right] at (0.8+2.7-0.6,-2.22-0.8) {$\tilde{\alpha}$};
\node[right] at (1.2+2.7-0.6,-2.20-0.8){$=B^{(L)}_{\tilde{\alpha}}$};
\end{tikzpicture}

\begin{tikzpicture}
\tenDotted{\mathcal{C}_j}{7}{-2}{1}{0}{0}{0}{0.8}
\node[right] at (0.8+5.4,-1.75) {$\alpha$};
\node[right] at (0.8+7,-1.75){$=\prod^N_{k=j+1}(g_j)^{-\frac{1}{2}}\big([A]^0_{\alpha,\alpha}\big)^{N-j}B^{(R)}_\alpha$};
\end{tikzpicture}

\noindent and we use the homogeneous limit of the MPO matrices introduced earlier
\begin{align}
A\equiv\lim_{\delta\to 0 }A^\pm\ .
\end{align}

\begin{figure}
    \centering
    \begin{tikzpicture}
    \node at (0,0) {\includegraphics[width=0.75\linewidth]{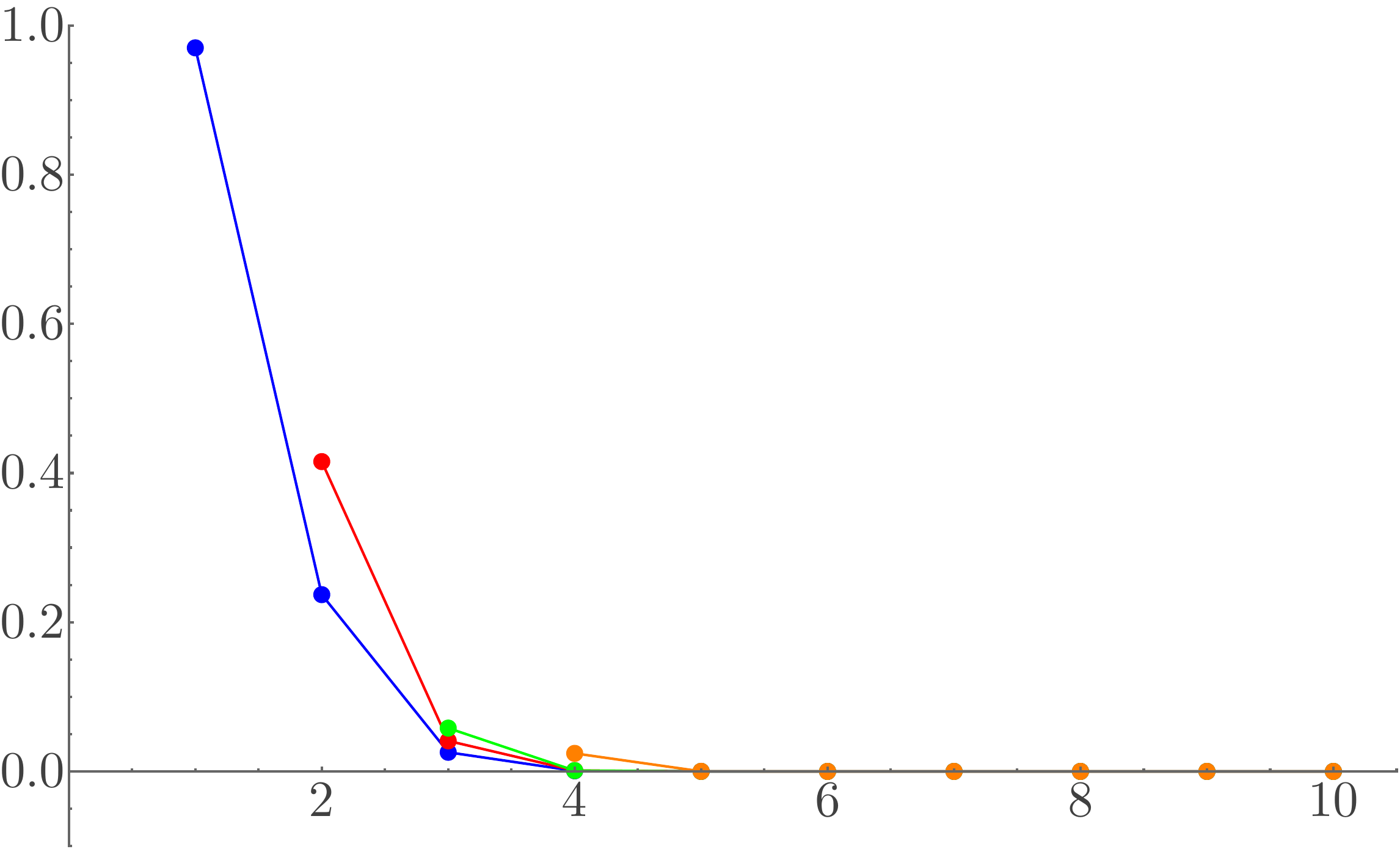}};
    \node at (6.15,-2.75) {$N$};
    \node at (-5.15,4) {$\frac{|\!|\Psi^{\rm ASEP}_j|\!|^2}{|\!|\scalebox{0.8}{\mathbbG{\Psi}}^{\rm ASEP}|\!|^2}$};

      \fill[blue]   (4,1.7+1) circle (1.825pt); \node[anchor=west] at (4+0.2,1.7+1) {$|\!|\Psi_1|\!|^2$};
      \fill[red]    (4,1.2+1) circle (1.825pt); \node[anchor=west] at (4+0.2,1.2+1) {$|\!|\Psi_2|\!|^2$};
      \fill[green]  (4,0.7+1) circle (1.825pt); \node[anchor=west] at (4+0.2,0.7+1) {$|\!|\Psi_3|\!|^2$};
      \fill[orange] (4,0.2+1) circle (1.825pt); \node[anchor=west] at (4+0.2,0.2+1) {$|\!|\Psi_4|\!|^2$};

    \end{tikzpicture}
\caption{Normalized Hilbert Schmidt norms of $\Psi^{\rm ASEP}_j$ for $j=1,2,3,4$ as functions of the system size $N$. We observe that the contribution of $\Psi_j^{\rm ASEP}$ becomes negligible as $N$ increases at fixed $j$, implying that that 
\fr{psiasep} is not localised in the vicinity of the boundary. The system parameters are $s^{(R)}=0.13$, $s^{(L)}=0.7$, $\varphi^{(L)}=0.68$, $\varphi^{(R)}=0.98$, $\xi^{(R)}=0.25$, $\eta=1.35$ and $\Lambda=1.1$. }
    \label{fig:ksfknjenfjdnfjdn}
\end{figure}

We find that $|\!|\Psi^{\rm ASEP}_j|\!|^2\to 0$ as $N\to \infty$ for any $j$ including $j=1$, yielding no localization at the boundary! This behaviour originates from the exponentially suppressing factor in the $\mathcal{C}_j$ matrix when $N\to \infty$, which was absent for the quantum circuit/XXZ case but needed in the ASEP to ensure \eqref{dssnjfngaöfnf}. Numerical data is represented in Figure \ref{fig:ksfknjenfjdnfjdn}.

%%%%%%%%%%%%%%%%%%%%%%%%%%%%%%%%%%%%%%%%%%%%%%%%%%%%%%%%%%%%%%%%%%%%%%%%%%%%%%%%%%%%%%%%%%%%%%%%%%%%%%%%%%%%%%%%%%%%%%%%%%%%%%%%%%%%%%%%%%%%%%%%%%%%%%%%%%%%%%%%%%%%%%%%%%%%%%%%%%%%%%%%
\section{Conclusions}
\label{sec:summary}
%%%%%%%%%%%%%%%%%%%
In this work, we have constructed exact strong zero mode operators for both integrable quantum brick-wall circuits and the spin-1/2 Heisenberg XXZ spin chain with the most general integrable boundary conditions compatible with respectively unitarity and Hermiticity. This generalizes previous studies in the literature \cite{fendley2016strong,vernier2024strong} in that the boundary conditions break the $U(1)$ symmetry present in the bulk of the system, and magnetization is therefore no longer conserved under time evolution. We have derived an explicit matrix product operator (MPO) form for the ESZM operators and used this representation to prove that they are localized (by construction) around the left boundary of the system.

The presence of an ESZM is expected to give rise to infinite coherence time for autocorrelation functions of certain operators localized near the boundary. We have verified numerically that this is indeed the case for $\sigma^z_1$ autocorrelation functions the XXZ model, and have observed that the asymptotic value reached at late time agrees with considerations based on the finite overlap of $\sigma^z_1$ with the ESZM operator.

There is a well-known map between the spin-1/2 Heisenberg XXZ chain with open boundary conditions and the asymmetric exclusion process and the existence of an ESZM operator in the former poses the question of what role, if any, this plays in the latter.
We have shown that under the map the ESZM loses its defining property of being spatially localized in the vicinity of the boundaries. This implies that it cannot affect the local properties of the ASEP in a significant way.

The effects of the presence of strong zero modes in kicked Ising models has been previously investigated in systems of transmon qubits in \cite{mi2022noise}. The brick-wall quantum circuit studied here has been realized in the same device \cite{morvan2022formation}, and it would be interesting to investigate boundary effects in this system in light of our findings.

%%%%%%%%%%%%%%%%%%%%%%%%%%%%%%%%%%%%%%%%%%%%%%%%%%%%%%%%%%%%%%%%%%%%%%%%%%%%%%%%%%%%%%%%%%%%%%%%%%%%%%%%%%%%%%%%%%%%%%%%%%%%%%%%%%%%%%%%%%%%%%%%%%%%%%%%%%%%%%%%%%%%%%%%%%%%%%%%%%%%%%%%

\section*{Acknowledgments}
This work was supported by the EPSRC under
grant EP/X030881/1. We are grateful to Eric Vernier and Paul Fendley for many helpful discussions, and Referee 1 for providing us with the argument based on von Neumann's mean ergodic theorem. 

%%%%%%%%%%%%%%%%%%%%%%%%%%%%%%%%%%%%%%%%%%%%%%%%%%%%%%%%%%%%%%%%%%%%%%%%%%%%%%%%%%%%%%%%%%%%%%%%%%%%%%%%%%%%%%%%%%%%%%%%%%%%%%%%%%%%%%%%%%%%%%%%%%%%%%%%%%%%%%%%%%%%%%%%%%%%%%%%%%%%%%%%

%%%%%%%%%%%%%%%%%%%%%%%%%%%%%%%%%%%%%%%%%%%%%%%%%%%%%%%%%%%%%%%%%%%%%%%%%%%%%%%%%%%%%%%%%%%%%%%%%%%%%%%%%%%%%%%%%%%%%%%%%%%%%%%%%%%%%%%%%%%%%%%%%%%%%%%%%%%%%%%%%%%%%%%%%%%%%%%%%%%%%%%%

\appendix

\section{Explicit matrix elements of the SZM as MPO\label{nsjdnsjdnsjdnjs}}
To get the homogeneous limit set $\delta=0$. The explicit non-vanishing matrix elements are given by 
\begin{align}
\big[A^{\pm}\big]^{0}_{0,0}=&\cos^2\big(\tfrac{\delta}{2} \big) \text{sech}\big(\tfrac{\ri\delta}{2} -\eta \big) \text{sech}\big(\tfrac{\ri\delta}{2} +\eta \big)\ ,\nn
\big[A^{\pm}\big]^{0}_{x,x}=&\cos^2\big(\tfrac{\delta}{2} \big)  \cosh (\eta ) \text{sech}\big(\tfrac{\ri\delta}{2} -\eta \big) \text{sech}\big(\tfrac{\ri\delta}{2} +\eta \big)\ ,\nn
\big[A^{\pm}\big]^{0}_{y,y}=&\cos^2\big(\tfrac{\delta}{2} \big)  \cosh (\eta ) \text{sech}\big(\tfrac{\ri\delta}{2} -\eta \big) \text{sech}\big(\tfrac{\ri\delta}{2} +\eta \big)\ ,\nn
\big[A^{\pm}\big]^{0}_{z,z}=&1\,,
\end{align}
\begin{align}
\big[A^{\pm}\big]^{x}_{0,x}=&+\big[A^{\pm}\big]^{y}_{0,y}=\ri \cos \big(\tfrac{\delta }{2}\big) \sinh (\eta ) \text{sech}\big(\tfrac{\ri\delta}{2} -\eta \big) \text{sech}\big(\tfrac{\ri\delta}{2} +\eta \big)\ ,\nn
\big[A^{\pm}\big]^{x}_{x,0}=&+\big[A^{\pm}\big]^{y}_{y,0}=\ri \cos \big(\tfrac{\delta }{2}\big) \sinh (\eta ) \cosh (\eta ) \text{sech}\big(\tfrac{\ri\delta}{2} -\eta \big) \text{sech}\big(\tfrac{\ri\delta}{2} +\eta \big)\ ,\nn
\big[A^{\pm}\big]^{x}_{y,z}=&-\big[A^{\pm}\big]^{y}_{x,z}=\mp\ri\sin \big(\tfrac{\delta}{2} \big) \sinh ^2(\eta ) \text{sech}\big(\tfrac{\ri\delta}{2} -\eta \big) \text{sech}\big(\tfrac{\ri\delta}{2} +\eta \big)\ ,
\end{align}
\begin{align}
\big[A^{\pm}\big]^{z}_{0,z}=&\sinh ^2(\eta ) \text{sech}\big(\tfrac{\ri\delta}{2} -\eta \big) \text{sech}\big(\tfrac{\ri\delta}{2} +\eta \big)\ ,\nn
\big[A^{\pm}\big]^{z}_{x,y}=&\pm \tfrac{1}{2}\sin(\delta) \sinh (\eta ) \text{sech}\big(\tfrac{\ri\delta}{2} -\eta \big) \text{sech}\big(\tfrac{\ri\delta}{2} +\eta \big)\ ,\nn
\big[A^{\pm}\big]^{z}_{y,x}=&\mp \tfrac{1}{2}\sin(\delta)  \sinh (\eta ) \text{sech}\big(\tfrac{\ri\delta}{2} -\eta \big) \text{sech}\big(\tfrac{\ri\delta}{2} +\eta \big)\ ,
\end{align}
\begin{align}
B^{(L)}_0=&1\ ,\nn
B^{(L)}_x=&-2\ri s^{(L)}\sin(\varphi^{(L)})\ ,\nn
B^{(L)}_y=&+2\ri s^{(L)}\cos(\varphi^{(L)})\ ,\nn
\end{align}
\begin{align}
B^{(R)}_0=&-2\ri\sinh(\eta)\sinh(\xi^{(R)})\ ,\nn
B^{(R)}_x=&-2 s^{(R)}\cos(\varphi^{(R)})\sinh(2\eta)\ ,\nn
B^{(R)}_y=&-2 s^{(R)}\sin(\varphi^{(R)})\sinh(2\eta)\ ,\nn
B^{(R)}_z=&-2\ri\cosh(\xi^{(R)})\cosh(\eta)\ .
\end{align}
%%%%%%%%%%%%%%%%%%%%%%%%%%%%%%%%%%%%%%%%%%%%%%%%%%%%%%%%%%%%%%%%%%%%%%%%%%%%%%%%%%%%%%%%%%%%%%%%%%%%%%%%%%%%%%%%%%%%%%%%%%%%%%%%%%%%%%%%%%%%%%%%%%%%%%%%%%%%%%%%%%%%%%%%%%%%%%%%%%%%%%%%

\section{Normalisation Constant for Finite system sizes\label{Normalisations}}
For completeness sake we here report the normalization constant on the ESZM in a finite volume $N$
\begin{align*}
\mathcal{N}=&\frac{2}{\kappa} \Bigg(\sinh ^2(2 \eta ) \Big[-4 \kappa \big(s^{(L)}\big)^2 \lambda _4^L \big(s^{(R)}\big)^2 \cos \big(2 (\phi^{(L)} -\phi^{(R)})\big)\\
%-----------------------------
&+2 \kappa \big(s^{(L)}\big)^2  \big(s^{(R)}\big)^2(\lambda _6^L+\lambda _7^L) -(\lambda _6^L-\lambda _7^L) \Big(\big(s^{(L)}\big)^2 \cosh \big(2 \xi^{(R)}\big)\\
%-----------------------------
&+2 \Big(\big(s^{(L)}\big)^2+1\Big) \big(s^{(R)}\big)^2-\big(s^{(L)}\big)^2\Big)\Big]\\
%-----------------------------
&-16 \sinh ^2(\eta ) \cosh (\eta ) \kappa s^{(L)}s^{(R)} \lambda _5^L  \sinh \big(\xi^{(R)}\big) \sin \big(\phi^{(L)}-\phi^{(R)}\big)\\
%-----------------------------
&-\cosh ^2(\eta ) \cosh ^2\big(\xi^{(R)}\big) \Big[-4 \cosh (2 \eta ) \big(s^{(L)}\big)^2 \Big(\lambda _6^L-\lambda _7^L\Big)\\
%-----------------------------
&+\kappa \Big(4 \big(s^{(L)}\big)^2 \Big(-2 \lambda _3^L+\lambda _6^L+\lambda _7^L\Big)+\lambda _6^L+\lambda _7^L\Big)-\Big(8 \big(s^{(L)}\big)^2+3\Big) \Big(\lambda _6^L-\lambda _7^L\Big)-2 \kappa \lambda _3^L\Big]\\
%-----------------------------
&+\sinh ^2(\eta ) \sinh ^2\big(\xi^{(R)}\big) \Big(\Big(\kappa+1\Big) \lambda _6^L+\Big(\kappa-1\Big) \lambda _7^L\Big)\Bigg)\,,
\end{align*}
where we recall $N=2L$ and we have further 
\begin{align*}
\kappa=&\sqrt{4 \cosh (2 \eta )+5}\\
\lambda_1=&\frac{(\cos\delta+1)^2}{\bigl(\cos\delta+\cosh(2\eta)\bigr)^2},\\
\lambda_2=&\frac{4\cos^4\!\bigl(\tfrac{\delta}{2}\bigr)\,\cosh^2\eta}{\bigl(\cos\delta+\cosh(2\eta)\bigr)^2},\\
\lambda_3=&1,\\
\lambda_4=&\frac{4\cos^4\!\bigl(\tfrac{\delta}{2}\bigr)\,\bigl(\cos\delta\,\cosh(2\eta)+1\bigr)^2}
{\bigl(\cos\delta+\cosh(2\eta)\bigr)^4},\\
\lambda_5=&\frac{4\cos^4\!\bigl(\tfrac{\delta}{2}\bigr)\,\cosh^2\eta\,\bigl(\cos\delta-\cosh(2\eta)+2\bigr)^2}
{\bigl(\cos\delta+\cosh(2\eta)\bigr)^4},\\
\lambda_6=&\frac{\cos^4\!\bigl(\tfrac{\delta}{2}\bigr)}{\kappa\,\bigl(\cos\delta+\cosh(2\eta)\bigr)^4}\!\left[
-20\,(2\cos\delta+1)\,\sinh^2\eta
+8\cos\delta\,\kappa\,\cosh^2\eta
+(2\cos(2\delta)+7)\,\kappa\right.\\
&\left.\quad-\cosh(2\eta)\Bigl(4\sinh^2\eta\,\bigl(8\cos\delta+4\cosh(2\eta)+9\bigr)+\kappa\Bigr)
+\bigl(\cosh(6\eta)-\cosh(4\eta)\bigr)\,\kappa
\right],\\
\lambda_7=&\frac{\cos^4\!\bigl(\tfrac{\delta}{2}\bigr)}{\kappa\,\bigl(\cos\delta+\cosh(2\eta)\bigr)^4}\!\left[
\;20\,(2\cos\delta+1)\,\sinh^2\eta
+8\cos\delta\,\kappa\,\cosh^2\eta
+(2\cos(2\delta)+7)\,\kappa\right.\\
&\left.\quad+\cosh(2\eta)\Bigl(4\sinh^2\eta\,\bigl(8\cos\delta+4\cosh(2\eta)+9\bigr)-\kappa\Bigr)
+\bigl(\cosh(6\eta)-\cosh(4\eta)\bigr)\,\kappa
\right].\\
\end{align*}
All the eigenvalues $\lambda_j$ are smaller than one expect $\lambda_3\equiv 1$, which the dominant term as $N\to \infty$.

\section{Diagonalisation of $\tilde{\mathcal{A}}$\label{sakdmskdmskd}}
We have 
\begin{align}
\tilde{\mathcal{A}}=SDS^{-1}
\end{align}
\begin{equation}\label{sdksndjsndjsndj}
\begin{aligned}
D&=\text{diag}\left\{d_1,d_2,d_2,d_3,d_3,d_4,d_4,d_5,d_6\right\}\\[4pt]
d_1 &= \frac{\cos\delta + 1}{\cos\delta + \cosh(2\eta)},\\[4pt]
d_2 &= \frac{2\cos^2\!\left(\frac{\delta}{2}\right)\big(\cos\delta\,\cosh(2\eta)+1\big)}{\big(\cos\delta+\cosh(2\eta)\big)^2},\\[4pt]
d_3 &= \frac{2\cos^2\!\left(\frac{\delta}{2}\right)\cosh\eta\,\big(\cos\delta-\cosh(2\eta)+2\big)}{\big(\cos\delta+\cosh(2\eta)\big)^2},\\[4pt]
d_4 &= \frac{\big(\cos\delta+1\big)\cosh\eta}{\cos\delta+\cosh(2\eta)},\\[4pt]
d_5 &= \frac{\cos^2\!\left(\frac{\delta}{2}\right)\Big(2\cos\delta-\cosh(2\eta)\,(\kappa-1)+\kappa+1\Big)}{\big(\cos\delta+\cosh(2\eta)\big)^2},\\[4pt]
d_6 &= \frac{\cos^2\!\left(\frac{\delta}{2}\right)\Big(2\cos\delta-\kappa+\cosh(2\eta)\,(\kappa+1)+1\Big)}{\big(\cos\delta+\cosh(2\eta)\big)^2}.
\end{aligned}
\end{equation}
\begin{align}
S=\left(
\begin{array}{ccccccccc}
0 & 0 & 0 & 0 & 0 & 0 & 0 & -\tfrac{1}{2}(\kappa+1)\,\text{sech}^2(\eta) & \tfrac{1}{2}(\kappa-1)\,\text{sech}^2(\eta) \\[4pt]
0 & 0 & 0 & 0 & -1 & 0 & 1 & 0 & 0 \\[4pt]
0 & 0 & 0 & -1 & 0 & 1 & 0 & 0 & 0 \\[4pt]
0 & 0 & 0 & 0 & 1 & 0 & 1 & 0 & 0 \\[4pt]
0 & -1 & 0 & 0 & 0 & 0 & 0 & 1 & 1 \\[4pt]
-1 & 0 & 1 & 0 & 0 & 0 & 0 & 0 & 0 \\[4pt]
0 & 0 & 0 & 1 & 0 & 1 & 0 & 0 & 0 \\[4pt]
1 & 0 & 1 & 0 & 0 & 0 & 0 & 0 & 0 \\[4pt]
0 & 1 & 0 & 0 & 0 & 0 & 0 & 1 & 1
\end{array}
\right)
\end{align}

\bibliography{bib}
\end{document}